\documentclass[nonacm,screen]{acmart}
\usepackage{booktabs}
\usepackage{tabularray}
\usepackage{tblr-extras}
\UseTblrLibrary{caption}
\usepackage{standalone}
\usepackage{tikz}
\usepackage{pgfplots}
\pgfplotsset{compat=1.18}
\usepackage{tcolorbox}

\begin{document}

\title{The Hidden Methodological Layer: Uncovering Threats to Validity in LLM-Assisted Empirical Software Engineering}

\author{Victória Gomes}
\affiliation{%
  \institution{Virginia Commonwealth University}
  \city{Richmond}
  \country{USA}
}
\email{oliveiragov@vcu.edu}

\author{Fabio Palomba}
\affiliation{%
  \institution{University of Salerno}
  \city{Fisciano}
  \country{Italy}}
\email{fpalomba@unisa.it}

\author{Robert Feldt}
\affiliation{%
  \institution{Chalmers University of Technology}
  \city{Gothenburg}
  \country{Sweden}}
\email{robert.feldt@chalmers.se}

\author{Thomas Zimmermann}
\affiliation{%
  \institution{University of California}
  \city{Oakland}
  \country{USA}}
\email{tzimmer@uci.edu}

\author{David Lo}
\affiliation{%
  \institution{Singapore Management University}
  \city{Singapore}
  \country{Singapore}}
\email{davidlo@smu.edu.sg}

\author{Rodrigo Spinola}
\affiliation{%
  \institution{Virginia Commonwealth University}
  \city{Richmond}
  \country{USA}}
\email{spinolaro@vcu.edu}

\renewcommand{\shortauthors}{Gomes et al.}
\renewcommand{\shorttitle}{The Hidden Methodological Layer}

\begin{abstract}
\textbf{Context:} Large Language Models (LLMs) are increasingly embedded in empirical software engineering (ESE) workflows, supporting tasks such as classification, coding, extraction, synthesis, and evaluation. Their probabilistic and opaque behavior, however, introduces methodological concerns that may affect empirical rigor.
\textbf{Aims:} This paper investigates threats to validity introduced by LLM-assisted empirical workflows and examines how LLM interactions influence empirical studies.
\textbf{Method:} We conducted a secondary analysis of 50 primary studies comprising 69 LLM-assisted research tasks identified through a systematic literature review. We examined the extent to which LLM-related threats were reported, how these threats were characterized, and how they related to established validity dimensions.
\textbf{Results:} LLM-related threats remain substantially underreported. We identified recurring threats including classification errors, prompt-dependent interpretation, workflow instability, evaluator bias, downstream propagation of outputs, and model dependence. These threats can affect multiple validity dimensions depending on where and how LLM outputs enter the workflow.
\textbf{Conclusion:} We introduce the concept of hidden methodological layer as a conceptual lens for understanding how LLMs influence empirical software engineering workflows. We argue that rigorous LLM-assisted ESE requires making prompts, model configurations, interaction protocols, validation procedures, and downstream reuse of LLM-generated outputs visible as part of the research method.
\end{abstract}

\ccsdesc[500]{General and reference~Surveys and overviews}

\keywords{Large language model, LLM, empirical software engineering, threats to validity}

\maketitle

\section{Introduction}
\label{sec:introduction}

Empirical Software Engineering (ESE) relies on carefully designed methodological processes to produce valid and reliable knowledge about software systems and practices \cite{Kitchenham2002preliminary, Runeson2009guidelines, ExperimentationSEWohlin}. This reliance on methodological rigor makes the explicit control and reporting of research decisions central to ESE, particularly those concerning data collection, construct operationalization, analysis procedures, and interpretation of results \cite{Easterbrook2008, Jedlitschka2008}. Together, these elements form the backbone of empirical rigor, enabling researchers to assess validity, replicate studies, and build cumulative knowledge \cite{Basili1999}.

Given ESE's dependence on the transparency and control of these methodological decisions, the increasing adoption of Large Language Models (LLMs) in software engineering introduces a fundamental shift in how empirical research is conducted \cite{Hou2023}. Unlike traditional tools, which are typically deterministic and explicitly configured, LLMs act as adaptive, general-purpose agents capable of interpreting instructions, generating outputs, and making implicit decisions during task execution \cite{Bommasani2021, Brown2020language, Wei2022chain}. As a result, they are no longer merely supporting tools, but active components in the research process, influencing how data is processed, interpreted, and transformed \cite{Zhang2025}.

This shift changes the methodological conditions under which empirical research is conducted. When researchers rely on LLMs to perform tasks such as classification, coding, or synthesis, part of the research logic becomes embedded in interactions with the model, including prompts, parameter settings, and iterative refinements \cite{Muller2023genaiqual}. These interactions may introduce assumptions, biases, and stochastic behaviors that mediate how empirical evidence is constructed, interpreted, and reported \cite{Sallou2024breaking, Weidinger2022}. Importantly, while these interactions provide some level of control, the internal behavior of LLMs remains largely opaque and only partially controllable, limiting the researcher’s ability to fully understand or predict how outputs are produced \cite{Bommasani2021, Weidinger2022}. Consequently, part of the methodological process becomes effectively hidden, operating beneath the observable workflow \cite{demoraisleca2025apps}.

While this hidden layer may streamline and scale empirical activities, it also raises critical concerns. If key methodological decisions are implicit, difficult to control, or inconsistently reported, then traditional assumptions about validity, reproducibility, and transparency may no longer hold. Understanding this shift is essential for ensuring that the integration of LLMs strengthens rather than undermines empirical rigor.

\subsection{Problem Statement}

Current research on LLM-assisted empirical software engineering has primarily focused on the capabilities and applications of these models, examining how they can support specific tasks or improve efficiency \cite{Hou2023}. However, much less attention has been given to the methodological implications of their use, particularly regarding how LLMs alter the structure and properties of empirical workflows \cite{Sallou2024breaking, Wagner2025guidelines}.

In this paper, we argue that this gap reflects a deeper methodological issue. Although LLMs may be used as analytical tools, their integration into empirical workflows can become methodologically consequential when model-mediated outputs participate in the construction, transformation, interpretation, or evaluation of empirical evidence. We refer to the resulting set of partially visible methodological influences as the \textbf{hidden methodological layer}. This layer is “hidden” not because the use of the LLM itself is concealed, but because part of its methodological influence is mediated through opaque, probabilistic, and input-sensitive model behavior, as well as through prompts, configurations, and interaction patterns that may be only partially inspectable or insufficiently reported. As a result, it becomes difficult not only to assess how results were produced and reproduce findings, but also to determine whether the methodological risks introduced by LLM use were explicitly recognized and addressed by researchers. Empirical studies may therefore be subject to validity concerns that arise not solely from explicit study design decisions, but also from how LLM-mediated processes operate within the broader empirical workflow, affecting established study-specific validity dimensions in subtle and sometimes difficult-to-observe ways

In practice, this means that constructs may be implicitly redefined through prompts, causal relationships may be confounded by stochastic model behavior, findings may depend on specific models or configurations, and conclusions may vary across repeated executions of the same workflow \cite{Gilardi2023, Wei2022chain}. Despite their importance, these issues have not yet been systematically conceptualized or analyzed.

\subsection{Objectives}

The goal of this paper is to characterize and conceptualize the hidden methodological layer introduced by LLMs in empirical software engineering, through the analysis of how threats to validity emerging from LLM-assisted workflows are reported, manifested, and related to established validity dimensions. To operationalize this goal, we investigate three complementary aspects: (i) the extent to which such threats are explicitly reported in the literature, (ii) the types of threats reported in LLM-assisted empirical workflows, and (iii) how these threats relate to classical validity dimensions in empirical software engineering. These objectives are organized around the following overarching research question:

\begin{tcolorbox}[title=Overarching Research Question, colframe=black, fontupper=\normalsize, fonttitle=\normalsize\bfseries]
How does the hidden methodological layer manifest in LLM-assisted empirical software engineering?
\end{tcolorbox}

The specific research questions derived from this overarching question are presented in Section~\ref{sec:method}, together with the methodological procedures adopted to investigate them.

\subsection{Study Overview}

This study builds upon a systematic literature review on LLM-assisted empirical software engineering, which analyzed 50 primary studies published between 2020 and 2025 across leading software engineering venues~\cite{gomes2026llmassistedESE}. The original SLR focused on characterizing the ESE research tasks supported by LLMs, the phases of the empirical research process in which they are applied, how they are integrated into empirical workflows, the benefits and limitations reported in the literature, and the extent to which studies report information required for reproducible LLM usage.

In contrast, the present work adopts a complementary perspective by performing a secondary analysis of the same corpus to investigate the methodological implications of using LLMs as part of empirical research tasks. Specifically, we focus on examining (i) whether threats to validity introduced by LLM-assisted research tasks are explicitly reported, (ii) what kinds of validity threats are reported as arising from the integration of LLMs into empirical workflows, and (iii) how these threats relate to classical validity dimensions in empirical software engineering.

To support these analyses, we introduce a conceptual model that represents LLM-assisted empirical workflows as a multi-layered structure. The model distinguishes between the observable elements of empirical processes, the interactions with LLMs, and the underlying mechanisms through which these interactions influence research outcomes, as well as their implications for threats to validity. This layered perspective provides a structured lens to systematically reason about where and how validity issues may arise.

\subsection{Contributions}

This paper makes the following contributions:

\begin{itemize}
    \item A conceptualization of the \textit{hidden methodological layer} introduced by LLMs in empirical software engineering workflows, together with a conceptual model describing the multi-layered structure of LLM-assisted empirical processes;
    \item An empirical investigation of how LLM-related threats to validity are currently reported and manifested across different forms of LLM involvement in empirical workflows;
    \item A taxonomy of reported LLM-related threats to validity, including their mapping to established study-specific validity dimensions, providing a structured framework for reasoning about methodological risks in LLM-assisted empirical software engineering.
\end{itemize}

\subsection{Paper Organization}

The remainder of this paper is organized as follows. Section~\ref{sec:background} presents the background and introduces key concepts related to empirical software engineering and LLM-assisted workflows. Section~\ref{sec:method} describes the research method, including the data source and the analysis procedure adopted in this study. Section~\ref{sec:results} presents the results, structured around the research questions. Section~\ref{sec:discussion} discusses the findings and introduces the proposed conceptual model, synthesizing how LLM-assisted workflows give rise to validity concerns. Section~\ref{sec:threats} outlines the threats to validity of this study.  Section~\ref{sec:conclusion} concludes the paper and outlines directions for future research.

\section{Background}
\label{sec:background}

In this section, we first revisit established perspectives on validity in empirical studies, highlighting the assumptions underlying traditional frameworks. We then characterize key properties of LLMs that are particularly relevant from a methodological perspective, providing the foundation for understanding how these systems influence empirical processes and contribute to validity concerns.

\subsection{Validity in Empirical Software Engineering}

Empirical software engineering relies on well-established frameworks to assess the rigor and credibility of research findings \cite{Kitchenham2002preliminary, Shull2008, Zhang2018empirical}. Among the most widely adopted are the validity dimensions articulated by \textit{Wohlin et al.}~\cite{ExperimentationSEWohlin}: construct validity, internal validity, external validity, and conclusion validity. These dimensions provide a structured lens for reasoning about whether a study accurately captures the intended concepts, establishes credible causal relationships, generalizes beyond the studied context, and supports statistically sound conclusions \cite{Basili1999, Runeson2009guidelines}.

Although these dimensions are central to empirical software engineering, validity concerns are not operationalized in the same way across all empirical methods. Different study types involve different sources of evidence, procedures, and inferential goals, and therefore emphasize different kinds of validity concerns. For example, experiments and case studies commonly rely on classical validity dimensions such as construct, internal, external, and conclusion validity, whereas secondary studies emphasize risks related to search, selection, data extraction, and synthesis. Survey-based research, in turn, foregrounds threats associated with sampling, non-response, instrument design, and response behavior. Table~\ref{tab:validity-threats} summarizes these validity dimensions and threat categories across empirical software engineering studies.

These established perspectives have been fundamental for understanding how validity threats arise from study design choices, data collection procedures, measurement decisions, and analysis methods \cite{Easterbrook2008, Jedlitschka2008}. However, they largely assume that empirical work is conducted either by human researchers or through tools whose behavior is deterministic and explicitly configured. This assumption becomes less sufficient when modern AI-based systems are integrated into research workflows, because such systems may participate in data transformation, interpretation, classification, synthesis, or evaluation in ways that are adaptive, probabilistic, and only partially transparent \cite{Bommasani2021, Shneiderman2020human}.

\begin{longtblr}[
caption = {Threats to validity across empirical software engineering study types.},
label = {tab:validity-threats}
]{
    colspec = {Q[c, m, 2cm] Q[c, m, 3cm] X[j, m]},
    hlines,
    row{1} = {bg=black, fg=white, font=\bfseries},
    hline{1,2,5} = {0.5pt, solid},
    rows = {font=\small},
    rowsep = 1pt
}
    \textbf{Study Type} & \textbf{Validity dimension} & \textbf{Description} \\
    
    \SetCell[r=4]{l}  Experiments, Case Studies, and MSR \cite{ExperimentationSEWohlin}
    & Construct Validity 
    & Extent to which the operationalization of variables accurately reflects the intended theoretical constructs. Threats arise from mismatches between concepts and their measurements.  \\
    & Internal Validity 
    & Degree to which observed effects can be attributed to the variables under study rather than confounding factors. Threats include biases and uncontrolled variables.   \\
    & External Validity 
    & Extent to which findings generalize beyond the study context. Threats occur when results do not hold across different populations, settings, or conditions.  \\
    & Conclusion Validity 
    & Reliability of conclusions drawn from the data, especially regarding statistical relationships. Threats include low statistical power and improper analysis methods.  \\
    
    \SetCell[r=4]{l} SLRs and Mapping Studies \cite{kitchenham2007guidelines}
    & Search Bias 
    & Incomplete or inadequate search strategies that fail to identify all relevant studies, potentially leading to skewed evidence.  \\
    & Selection Bias 
    & Incorrect inclusion or exclusion of studies due to ambiguous or inconsistently applied selection criteria.  \\
    & Data Extraction Bias 
    & Errors or inconsistencies in extracting data from primary studies, which may affect synthesis results.  \\
    & Synthesis Bias 
    & Inappropriate aggregation or interpretation of results across studies, potentially leading to misleading conclusions.  \\
    
    \SetCell[r=4]{l} Surveys \cite{linaker2015guidelines}
    & Sampling Bias 
    & The selected sample is not representative of the target population, limiting generalizability.  \\
    & Non-response Bias 
    & Differences between respondents and non-respondents that may distort results.  \\
    & Instrument Bias 
    & Poorly designed survey instruments (e.g., ambiguous or leading questions) that affect the accuracy of responses. \\
    & Response Bias 
    & Systematic tendencies of respondents to answer inaccurately or untruthfully (e.g., social desirability bias). \\
\end{longtblr}

\subsection{Methodologically Relevant Properties of LLMs}
\label{sec:llm-characteristics}

The methodological impact of LLMs on empirical research is closely tied to a set of properties that distinguish them from traditional deterministic tools. These properties shape how LLMs influence empirical workflows and contribute to validity concerns in LLM-assisted studies. Prior research on LLMs and neural text generation has extensively discussed these characteristics~\cite{bender2021stochastic, Bommasani2021,henderson2018deep,ji2023survey,mehrabi2021survey,openai2024gpt4,reynolds2021prompt,zhao2021calibrate}:

\begin{itemize}
    \item \textbf{Opacity.} The internal reasoning processes of LLMs are not fully observable or explainable. While models can generate outputs and, in some cases, provide explanations, the underlying decision-making process remains largely inaccessible, making it difficult to inspect or justify intermediate analytical steps.

    \item \textbf{Probabilistic Outputs.} LLMs operate as probabilistic systems, meaning that identical inputs may yield different outputs across executions. This non-determinism introduces variability that can affect the stability, reproducibility, and comparability of empirical results across runs and settings.

    \item \textbf{Input Sensitivity.} Model outputs are highly sensitive to prompt formulation. Small changes in wording, structure, or context may lead to substantially different responses, effectively turning prompt design into a critical and often implicit methodological decision.

    \item \textbf{Hidden Configuration.} The behavior of LLMs is influenced by factors such as model version, parameter settings (e.g., temperature), and system-level prompts. These elements are frequently underreported, yet they can significantly affect outputs and, consequently, research findings.

    \item \textbf{Autonomous Influence.} LLMs can directly shape research outcomes by generating labels, classifications, or interpretations that may be incorporated into the analysis without systematic human validation, particularly in large-scale or automated workflows.

    \item \textbf{Training Data Dependence.} LLM outputs reflect patterns learned from large and heterogeneous training corpora. As a result, they may encode biases, outdated information, or domain-specific gaps that can influence the interpretation and generation of empirical data.

    \item \textbf{Contextual Variability.} Outputs may depend on broader interaction context, including prior prompts or conversation history. This makes results sensitive not only to the immediate input but also to the sequence of interactions leading to it.

    \item \textbf{Output Plausibility.} LLMs are optimized to produce fluent and contextually plausible outputs, which may appear correct even when they are factually inaccurate or unsupported by evidence. This tendency can make errors difficult to detect, particularly in tasks involving classification, summarization, or synthesis.
\end{itemize}

These properties expose a fundamental limitation of existing validity frameworks: they do not explicitly account for the methodological implications of LLM-assisted processes. While prior work provides a solid foundation for reasoning about threats arising from study design and data collection, it offers limited guidance on how to account for variability, bias, and hidden decision points introduced by LLMs. As a result, important aspects of the empirical process may remain implicit, hindering transparency, reproducibility, and critical assessment of results. Addressing this gap requires a systematic examination of how LLMs are integrated into empirical workflows and how their properties shape research outcomes. In the following section, we describe the methodological approach adopted in this study to investigate these issues and to characterize \textit{threats to validity introduced by LLM-assisted research tasks}.

\section{Research Method}
\label{sec:method}

This section describes the methodological approach adopted in this study. Building on a previously conducted systematic literature review \cite{gomes2026llmassistedESE}, we extend the original study by introducing additional data collection and analysis procedures to address the new research questions investigated in this paper. While the underlying review protocol remains unchanged, the present work focuses on extracting and analyzing information related to the threats to validity that arise in LLM-assisted empirical software engineering. The following subsections detail the reuse of the base SLR protocol, and the specific procedures employed for data collection and analysis.

\subsection{Research Questions}

To operationalize the overarching research question introduced in Section~\ref{sec:introduction}, we investigate three complementary aspects of LLM-related threats to validity: the extent to which they are explicitly reported, the types of threats reported, and how these threats relate to established validity dimensions. These aspects are captured by the research questions presented in Table~\ref{tab:research-questions}.

\begin{table}[ht]
\caption{Specific research questions addressed in the study.}
\label{tab:research-questions}
\begin{tblr} {
    colspec = {Q[c, m, 1cm] X[j, m]},
    hlines,
    row{1} = {bg=black, fg=white, font=\bfseries},
    hline{5} = {1.5pt, solid},
    rows = {font=\small},
    rowsep = 1pt
}
    \textbf{Id} & \textbf{Research question} \\
    
    RQ1 & To what extent are threats to validity introduced by LLM-assisted research tasks explicitly reported in empirical software engineering workflows? \\
    
    RQ2 & What threats to validity introduced by LLM-assisted research tasks are reported in empirical software engineering workflows? \\
    
    RQ3 & How do reported threats to validity introduced by LLM-assisted research tasks relate to established validity dimensions in empirical software engineering? \\
\end{tblr}
\end{table}

The first aspect concerns the visibility of LLM-related threats in the literature. As LLMs are increasingly used to support or automate empirical research activities, their integration into research workflows introduces methodological risks that may not always be explicitly acknowledged. Motivated by this concern, \textbf{RQ1} investigates the extent to which threats to validity introduced by LLM-assisted research tasks are explicitly reported, and whether this reporting varies according to the role assigned to the LLM within the empirical workflow. This focus on reporting is important because the absence of explicit discussion can make it difficult to assess how LLM use may have influenced the research process, limiting transparency, reproducibility, and the critical evaluation of LLM-assisted workflows.

While RQ1 examines whether LLM-related threats are made explicit, it does not explain what kinds of methodological risks are being reported when such threats are acknowledged. \textbf{RQ2} addresses this next step by focusing on the nature of the reported threats. Characterizing these threats is necessary because LLM use can affect empirical workflows through different mechanisms, including classification errors, hallucinations, prompt sensitivity, workflow instability, evaluator bias, and dependence on specific models or configurations. This question therefore aims to identify and synthesize recurring forms of threats to validity associated with LLM-assisted research tasks, supporting the development of a taxonomy of reported methodological risks in LLM-assisted empirical workflows.

Finally, \textbf{RQ3} examines how the identified threats intersect with established validity dimensions in empirical software engineering. This is important because traditional validity frameworks were developed primarily around human-driven procedures or deterministic tools, whereas LLMs introduce probabilistic, opaque, and prompt-sensitive mechanisms into empirical workflows. As a result, LLM-related threats may cut across conventional validity categories rather than aligning with a single dimension. \textbf{RQ3} therefore investigates how reported threats associated with LLM-assisted research tasks are distributed across classical validity dimensions, clarifying how LLM use affects the assessment of empirical rigor.

\subsection{Base SLR Protocol}

This study builds on a previously conducted SLR on the use of LLMs in empirical software engineering \cite{gomes2026llmassistedESE}. While the original review characterized how LLMs are used in empirical software engineering, including their benefits and limitations, the present study extends that work by investigating their methodological implications, particularly the threats to validity introduced by LLM-assisted research.

The original SLR examined peer-reviewed studies published between 2020 and 2025. Because the use of LLMs to support empirical research is often described within methodological sections rather than titles or abstracts, the review adopted a curated venue-based search strategy instead of relying solely on database search strings. The search covered 12 leading software engineering conferences and journals (ICSE, FSE, EASE, MSR, ESEM, CAIN, SANER, EMSE, IST, JSS, TSE, and TOSEM), followed by full-text keyword searches using terms such as \emph{Generative AI}, \emph{LLM}, \emph{ChatGPT}, \emph{Copilot}, \emph{Gemini}, and \emph{LLaMA}. Candidate studies were then screened using predefined inclusion and exclusion criteria to identify papers in which LLMs supported or were empirically investigated within empirical software engineering research workflows. This process resulted in a final corpus of 50 primary studies.

In this study, we reuse the same corpus of primary studies and conduct an additional round of data extraction and analysis guided by the new research questions introduced in this paper. The complete SLR protocol, including the search strategy, venue selection, study selection process, and inclusion/exclusion criteria, is described in detail in our previous work \cite{gomes2026llmassistedESE}. By retaining the same study corpus and methodological foundations, we ensure consistency with the original review while extending the analysis to investigate threats to validity associated with LLM-assisted empirical research. An overview of the primary studies included in the analysis is presented in Section~\ref{subsec:primary-studies}.

\subsection{Data Extraction}

We collected one additional data item from each study: threats to validity related to LLM use. For this item, we extracted verbatim excerpts from the papers to preserve the authors’ original wording and support traceability throughout the analysis. We considered only threats explicitly reported by the study authors as related to LLM use, rather than threats inferred by us from the study design or the presence of LLM-assisted tasks. As a result, our analysis reflects how LLM-related threats are currently reported and framed in the literature. 

Importantly, the data extraction was conducted at the level of the LLM-assisted empirical research task, rather than at the paper level. In our previous SLR, we had already identified and coded the empirical research tasks supported by LLMs in each study. Building on that analysis, for the present study, we preserved the association between extracted threats and the corresponding LLM-assisted tasks as explicitly reported by the authors. Since studies typically described the specific activity in which a validity concern emerged, this enabled us to maintain the connection between reported threats and the particular use of LLMs within the empirical workflow.

This task-based perspective was particularly important because some studies used LLMs to support multiple empirical research tasks, while only acknowledging threats to validity for a subset of those uses. If the analysis had been conducted only at the paper level, such studies would be treated as having reported LLM-related validity threats overall, potentially obscuring the fact that other LLM-assisted tasks within the same study were not discussed from a validity perspective. Analyzing threats at the task level therefore allowed us to obtain a more precise view of how studies acknowledge the validity implications of LLM use across different forms of involvement in empirical research workflows.

\subsection{Data Analysis}

\begin{figure}
    \centering
    \includegraphics[width=\linewidth]{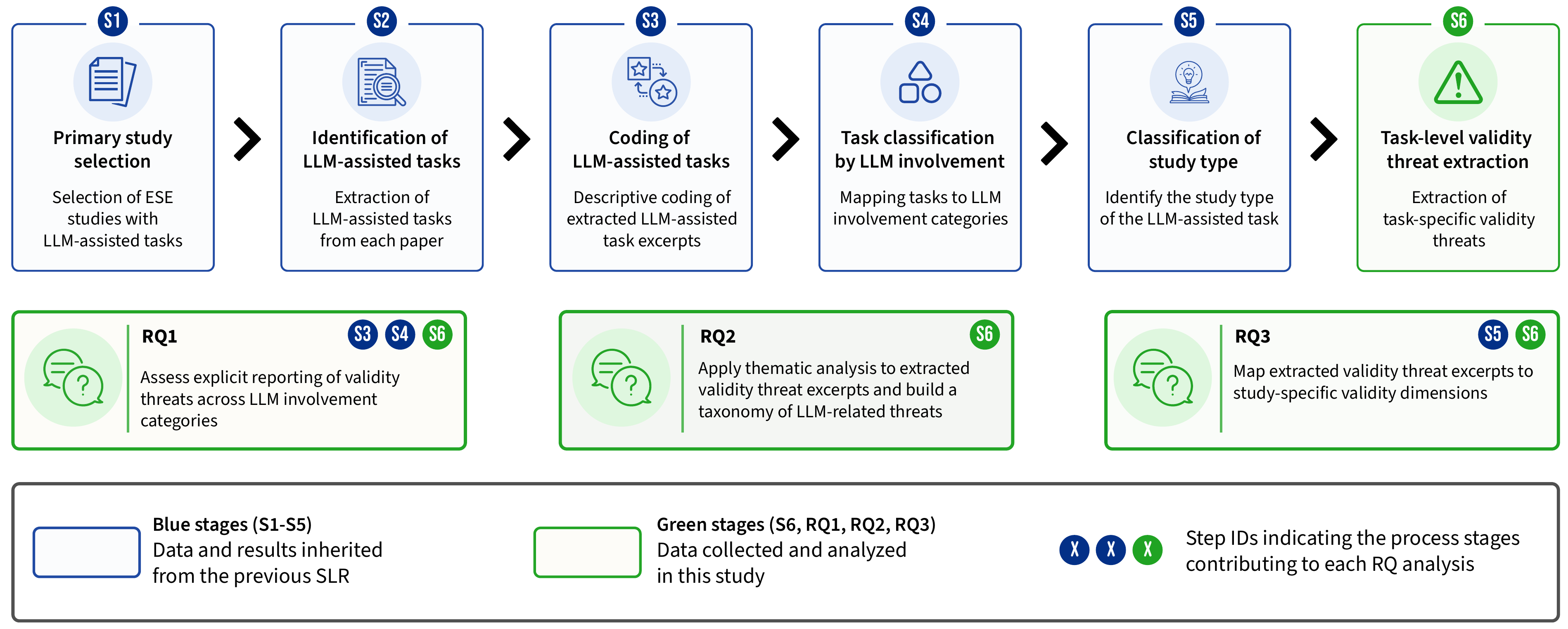}
    \caption{Overview of the study design and analytical procedure.}
    \Description{Six-stage workflow from primary-study selection to task-level validity-threat extraction, followed by analyses for three research questions. Stages S1–S5, shown in blue, reuse data and results from the previous SLR; S6 and RQ1–RQ3, shown in green, represent new data collection and analysis. RQ1 examines threats across LLM-involvement categories, RQ2 develops a taxonomy of LLM-related threats, and RQ3 maps threats to study-specific validity dimensions.}
    \label{fig:studyDesignAndAnalysis}
\end{figure}

The data analysis procedures were organized according to the three specific research questions. Figure~\ref{fig:studyDesignAndAnalysis} illustrates the overall methodological workflow adopted in this study, highlighting both the stages inherited from our previous SLR and the analyses newly conducted in the present work. The figure also shows how each research question draws upon specific process stages, making explicit the provenance of the data and analytical procedures used to address RQ1, RQ2, and RQ3.

To answer \textbf{RQ1 - To what extent are threats to validity introduced by LLM-assisted research tasks explicitly reported in empirical software engineering workflows?}, we examined how frequently threats to validity introduced by LLM-assisted research tasks were explicitly reported across the analyzed corpus. 

For each LLM-assisted task, we recorded whether the corresponding paper explicitly reported at least one threat to validity associated with LLM use. We then organized these results according to the taxonomy of LLM involvement established in our SLR: \textit{augmentation}, \textit{automation}, \textit{decision support}, and \textit{evaluation} \cite{gomes2026llmassistedESE}. In this taxonomy, \textit{augmentation} refers to scenarios in which LLMs support or accelerate human activities while the researcher remains primarily responsible for the task; \textit{automation} refers to workflows in which the LLM independently performs substantial parts of the empirical task with limited human intervention; \textit{decision support} captures cases in which the LLM provides recommendations, suggestions, or analytical insights that inform human decision making; and \textit{evaluation} refers to situations in which the LLM acts as an assessor or judge, for example by evaluating quality, correctness, similarity, or relevance. Organizing the analysis according to these categories allowed us to investigate whether the explicit reporting of LLM-related validity threats varied depending on the role assigned to the LLM within the empirical workflow.

To answer \textbf{RQ2 - What threats to validity introduced by LLM-assisted research tasks are reported in empirical software engineering workflows?}, we analyzed the reported threats through thematic analysis, following established guidelines for thematic synthesis in software engineering research \cite{ThematicAnalysisSE}. The analysis was conducted only for the subset of tasks in which at least one LLM-related validity threat was explicitly reported. We adopted an inductive coding process in which the extracted verbatim excerpts were first examined and assigned descriptive codes representing the specific validity concerns identified by the study authors. Conceptually related codes were then grouped into lower-level themes representing recurring forms of methodological risk introduced by LLM use. Finally, these lower-level themes were further abstracted into higher-level themes to characterize broader patterns in how threats to validity introduced by LLM-assisted research tasks are manifested and discussed in the literature.

The coding and synthesis process was conducted iteratively by one researcher and reviewed by a second researcher. We coded 33 extracted threat excerpts in total, treating each excerpt as a distinct threat occurrence. Because a single LLM-assisted task could be associated with more than one reported threat, multiple coded excerpts could be linked to the same task. The first researcher generated the initial codes, interpretations, and candidate themes, which were then reviewed and discussed with the second researcher across two refinement rounds. During these rounds, the researchers revisited the coded excerpts to improve conceptual consistency, reduce overlaps between categories, and ensure that each theme remained aligned with the methodological implications reported in the primary studies. Disagreements or ambiguous cases were resolved through discussion until consensus was reached. The resulting taxonomy therefore reflects iterative abstraction, peer review of coding decisions, and consensus-based refinement.

To answer \textbf{RQ3 - How do reported threats to validity introduced by LLM-assisted research tasks relate to established validity dimensions in empirical software engineering?}, we mapped each identified threat to its corresponding validity dimension. This mapping was conducted using the validity taxonomy appropriate for the type of empirical study, as presented in Table~\ref{tab:validity-threats}. Since the study type of each paper had already been extracted during our previous SLR, we used this information to determine the corresponding validity dimension before performing the classification. Each identified threat was then examined in terms of its underlying methodological implication and assigned to the validity dimension whose definition most closely reflected the nature of the concern being reported. For example, threats related to unreliable outputs were mapped to dimensions associated with construct validity, whereas concerns involving limited generalizability or contextual dependency were mapped to external validity dimensions. This procedure enabled us to systematically analyze how LLM-related validity threats manifest across different forms of empirical software engineering research and how they align with study-specific validity concerns.

The mapping process was conducted collaboratively, by one researcher and one reviewer, based on the definitions associated with each validity framework. Because some threats to validity introduced by LLM-assisted research tasks affected multiple aspects of empirical rigor simultaneously, a single threat could be associated with more than one validity dimension when conceptually justified by the reported methodological implications. The mapping decisions were iteratively reviewed and refined to improve consistency across the analyzed studies.

\section{Analysis of the Results}
\label{sec:results}

This section presents the results of the study, organized according to the specific research questions introduced in Section~\ref{sec:method}. Together, these analyses provide complementary perspectives on how the hidden methodological layer manifests in LLM-assisted empirical software engineering. We first present an overview of selected primary studies. Next, we examine the extent to which LLM-related validity threats are explicitly acknowledged in the literature (RQ1), providing a view of the visibility of these concerns across different forms of LLM involvement in empirical workflows. We then characterize the reported threats through the development of a taxonomy of recurring methodological risks (RQ2). Finally, we analyze how these threats relate to established validity dimensions in empirical software engineering (RQ3), providing a structured view of their implications for empirical rigor and research credibility.

To support transparency and traceability, the replication package for this study~\cite{replicationPackage} provides the data used in these analyses, including the extracted threat excerpts, coding decisions, threat categories, and validity-dimension mappings.

\subsection{Overview of the Primary Studies}
\label{subsec:primary-studies}

The analyzed corpus consisted of 50 primary studies published between 2020 and 2025, comprising 69 LLM-assisted empirical software engineering research tasks. When analyzed by task, the corpus covered multiple empirical research methodologies, including mining software repositories (34 tasks), controlled experiments and case studies (21 tasks), secondary studies such as systematic literature reviews and systematic mapping studies (11 tasks), and survey-based research (3 tasks). Across the corpus, LLMs were predominantly applied to structured and automation-oriented activities, particularly classification (17 tasks), screening (14 tasks), evaluation (11 tasks), synthetic data generation (8 tasks), content transformation (5 tasks), pattern analysis (5 tasks), and information extraction (5 tasks). 

The analyzed studies also revealed a strong concentration around proprietary conversational LLMs. Overall, the 50 papers reported 62 individual model usages, since several studies employed more than one model. ChatGPT alone accounted for 38 usages, followed by Llama (8), Claude (4), and DeepSeek (4). Only two studies did not explicitly report the adopted LLM model. In terms of publication trends, the analyzed corpus further indicates that LLM-assisted empirical software engineering remains an emerging methodological area, with the large majority of selected studies appearing after 2023, following the widespread adoption of ChatGPT and related conversational LLMs.

The diversity of empirical methods, research tasks, and forms of LLM involvement represented in the analyzed corpus provided a broad empirical basis for investigating the reporting, characterization, and validity implications of threats introduced by LLM-assisted research tasks across empirical software engineering workflows.

\subsection{RQ1: Reporting of LLM-related threats to validity}
\label{subsec:rq1}

Among the 69 LLM-assisted research tasks analyzed, 30 tasks (43.5\%) explicitly reported at least one LLM-related threat to validity. In contrast, the remaining 39 tasks (56.5\%) did not explicitly acknowledge or discuss potential validity concerns introduced by LLM use. 

\begin{figure}[ht]
    \centering   
    \definecolor{customDarkBlue}{HTML}{133C55}
\definecolor{customLightBlue}{HTML}{3DADFF}

\begin{tikzpicture}
    \begin{axis}[
        ybar,
        height=7cm,
        width=12cm,
        bar width=15pt,
        ylabel={Number of Tasks},
        xlabel={Nature of LLM involvement},
        xlabel style={yshift=-10pt},
        symbolic x coords={Augmentation, Automation, Decision Support, Evaluation},
        xtick=data,
        nodes near coords,
        nodes near coords style={font=\footnotesize, /pgf/number format/fixed},
        ymin=0, ymax=25,
        legend style={
            at={(0.36, 1.08)},
            anchor=south, 
            legend columns=-1, 
            draw=none,
            /tikz/every even column/.append style={column sep=15pt}
        },
        axis x line*=bottom,
        axis y line*=left,
        ymajorgrids=true,
        grid style={gray!20},
        xticklabel style={rotate=0, anchor=center, yshift=-8pt}
    ]
        \addplot[fill=customDarkBlue, draw=none] coordinates {
            (Augmentation, 10) (Automation, 20) (Decision Support, 1) (Evaluation, 8)
        };
        
        \addplot[fill=customLightBlue, draw=none] coordinates {
            (Augmentation, 3) (Automation, 22) (Decision Support, 0) (Evaluation, 5)
        };
        
        \legend{No Threats Reported, Threats reported}
    \end{axis}
\end{tikzpicture}
    \caption{Distribution of LLM-assisted tasks with and without explicitly reported LLM-related threats to validity, grouped by nature of LLM involvement.}
    \Description{Grouped bar chart showing reported validity threats across four types of LLM involvement. Augmentation has 10 tasks with no threats reported and 3 with threats; automation has 20 and 22, respectively; decision support has 1 and 0; and evaluation has 8 and 5. Automation has the highest task counts and is the only category where tasks reporting threats exceed those reporting no threats.}
    \label{fig:LLM-involvement-by-reported-vs-not-reported-threat-to-validity}
\end{figure}
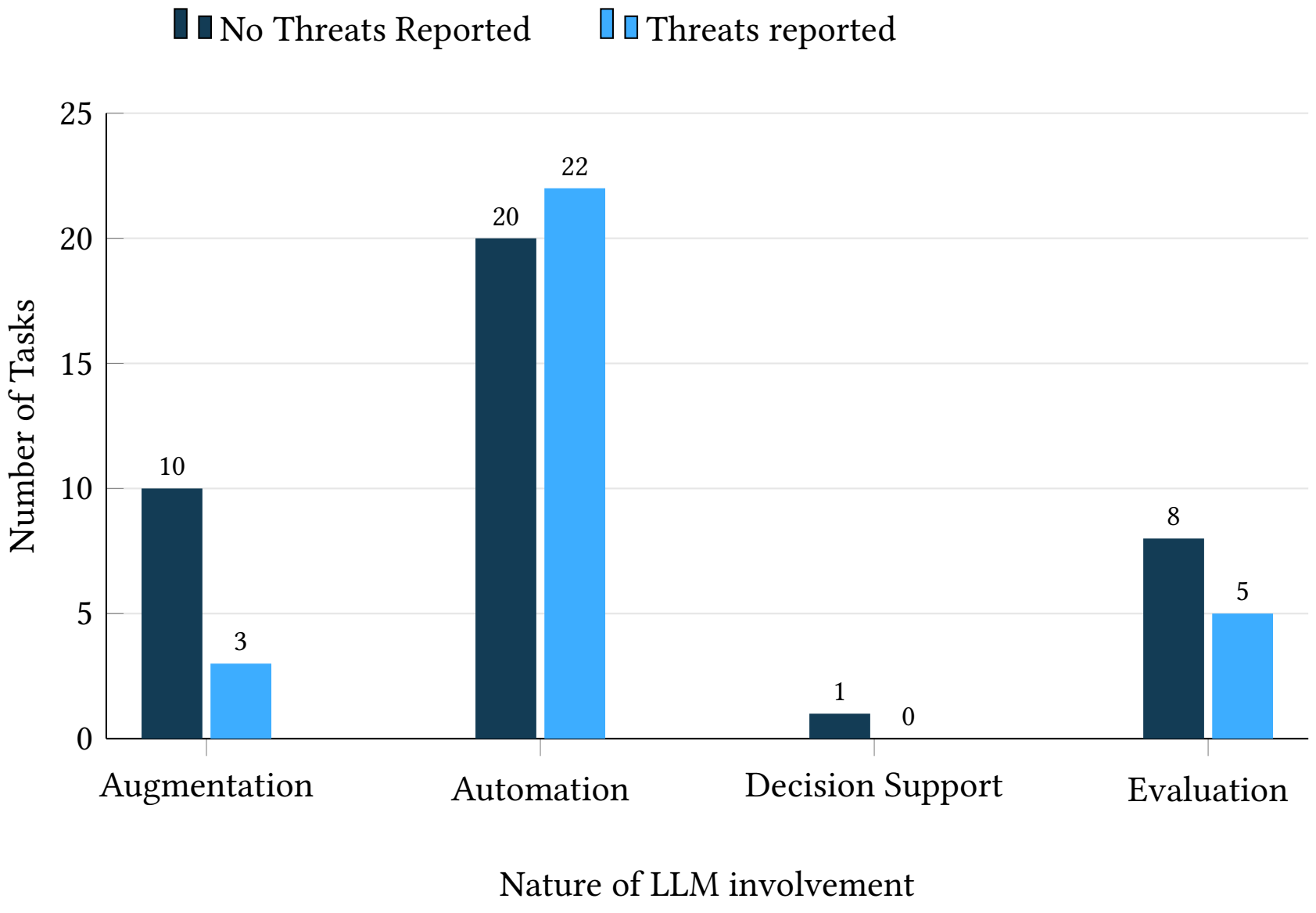

To further examine this reporting pattern, we analyzed whether the explicit acknowledgment of LLM-related threats varied according to the role assigned to the model in the empirical workflow. This distinction is important because the methodological implications of LLM use may depend on how the model is integrated into the research process. For example, risks associated with automated classification or screening may differ from those arising in assistive settings where researchers retain stronger oversight over interpretation and decision-making. Therefore, we organized the analyzed tasks according to the taxonomy of LLM involvement established in our previous work: \textit{augmentation}, \textit{automation}, \textit{decision support}, and \textit{evaluation}~\cite{gomes2026llmassistedESE}. Figure~\ref{fig:LLM-involvement-by-reported-vs-not-reported-threat-to-validity} presents, for each involvement type, the number of tasks that explicitly reported at least one LLM-related threat to validity.

The results indicate that explicit reporting of LLM-related validity threats differs according to how the LLM is positioned within the empirical workflow. This difference is particularly evident in tasks categorized as \textit{automation}, where the LLM performs substantial parts of the empirical task with limited human intervention. Among these tasks, 22 of 42 (52.4\%) explicitly reported at least one LLM-related threat to validity, making automation the category with the highest threat-reporting rate. This suggests that studies involving more autonomous forms of LLM use are more likely to acknowledge explicit methodological concerns.

At the same time, however, the \textit{automation} category also contained 20 tasks in which no LLM-related threat to validity was explicitly discussed. This result is particularly noteworthy because automation-oriented workflows are those in which LLM usage may introduce stronger methodological risks. When an LLM performs substantial parts of an empirical task with limited human intervention, the absence of explicit threat reporting becomes especially concerning, as it may obscure whether researchers assessed the reliability of generated outputs, controlled for hallucinations or misclassifications, evaluated prompt sensitivity, or validated the automated procedure itself. Therefore, the \textit{automation} category stands out not only as the form of involvement most frequently associated with reported threats, but also as a particularly sensitive context in which missing threat reporting and management may have greater methodological consequences.

In contrast, \textit{augmentation} tasks showed the lowest reporting rate, with only 3 of 13 tasks (23.1\%) acknowledging LLM-related threats. This may indicate that, when LLMs are used mainly to support or accelerate human work, authors are less likely to frame their use as introducing distinct validity concerns. In augmentation settings, the presence of human oversight may lead authors to perceive the LLM as less methodologically consequential. Nevertheless, this interpretation requires caution. Human involvement does not eliminate LLM-related risks; it may only change how they appear. For example, an LLM-generated suggestion, summary, or intermediate output may still influence how researchers interpret data, prioritize evidence, or make analytical decisions, even when the final judgment remains human.

Tasks categorized as \textit{evaluation} showed an intermediate pattern: 5 of 13 tasks (38.5\%) had at least one LLM-related threat reported in the corresponding paper, while 8 did not. This role deserves particular attention because LLMs used for evaluation are not merely producing intermediate artifacts; they are participating in judgment. When LLMs are used as evaluators, judges, or assessors, they can influence how quality, correctness, relevance, similarity, or performance are measured. Therefore, even though the reporting rate for evaluation was lower than automation, the methodological stakes remain high. In these cases, threats such as evaluator bias, inconsistency, misalignment with human judgment, and inaccurate assessment may directly affect the study's findings.

The only task categorized as \textit{decision support} did not report any LLM-related threat to validity. Consistent with the interpretation above, this should be understood as a reporting outcome rather than as evidence that decision-support uses are methodologically unproblematic. Conceptually, decision support may still introduce validity concerns because LLM-generated recommendations, rankings, explanations, or interpretations can shape human decisions even when the final choice remains with the researcher. Because this category contains only one task, however, the result does not support substantive comparison with the other forms of LLM involvement.

The high number of tasks without explicit threat reporting reveals an important transparency gap in LLM-assisted empirical software engineering research. When studies incorporate LLMs into research tasks but do not discuss their validity implications, readers are left with limited insight into how the use of these models may have shaped the research process. This absence of reporting does not indicate that LLM-related threats were absent; rather, it means that the methodological account does not provide enough information to assess whether such risks were considered, monitored, or controlled. As a result, readers cannot easily determine how LLM outputs were checked, how errors or biases may have affected the analysis, or whether safeguards were applied to address risks such as hallucinations, misclassification, prompt sensitivity, unstable outputs, or evaluator bias. This underreporting weakens the methodological traceability of LLM-assisted workflows and limits the extent to which their findings can be assessed, reproduced, or reused.

\subsection{RQ2: LLM-related threats to validity}

The previous analysis showed that many LLM-assisted tasks did not explicitly report LLM-related threats to validity. For the tasks in which threats were reported, however, the authors often reported more than one validity concern associated with LLM use, meaning that the number of identified threats exceeds the number of tasks in which threats were reported.

\begin{figure}[ht]
    \centering
    \includegraphics[width=0.9\linewidth]{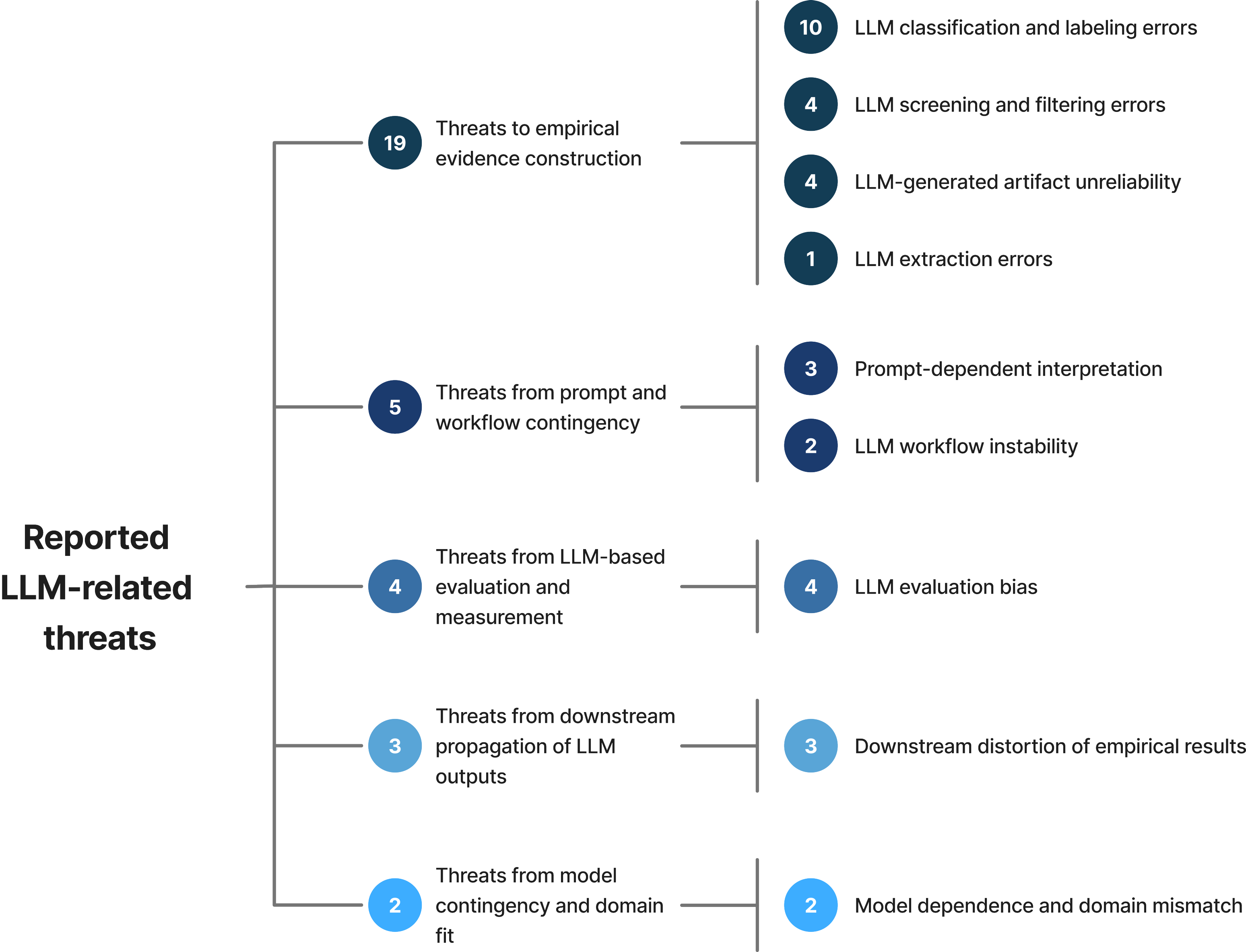}
    \caption{Taxonomy of reported LLM-related threats to validity, where numbers indicate coded threat occurrences extracted from verbatim threat excerpts.}
    \Description{Hierarchical diagram summarizing reported LLM-related threats across five threat categories. Threats to empirical evidence construction is the largest category, with 19 instances: 10 LLM classification and labeling errors, 4 LLM screening and filtering errors, 4 LLM-generated artifact unreliability, and 1 LLM extraction error. Threats from prompt and workflow contingency includes 5 instances: 3 prompt-dependent interpretation and 2 LLM workflow instability. Threats from LLM-based evaluation and measurement includes 4 instances, all involving LLM evaluation bias. Threats from downstream propagation of LLM outputs includes 3 instances, all involving downstream distortion of empirical results. Threats from model contingency and domain fit includes 2 instances, all involving model dependence and domain mismatch. Numbered blue circles indicate the frequency associated with each category and specific threat.}
    \label{fig:taxonomy-llm-threats}
\end{figure}

To characterize these concerns, we synthesized a taxonomy of LLM-related threats to validity. In this taxonomy, lower-level themes represent specific validity threats, while the higher-level themes group related threats according to broader forms of methodological risk introduced by LLM use. Figure~\ref{fig:taxonomy-llm-threats} provides a visual overview of the taxonomy and the distribution of coded threats across its different levels, while Table~\ref{tab:taxonomy-llm-threats} defines each threat and clarifies its methodological meaning.

The identified threats were distributed across five higher-level themes. Most threats were concentrated in \textit{threats to empirical evidence construction}, which accounted for 19 coded occurrences. This concentration indicates that LLM-related validity threats were more likely to be reported when the model was used to produce, transform, classify, filter, or extract the empirical evidence used in later analyses.

Within this higher-level theme, the most frequent lower-level theme was \textit{LLM classification and labeling errors}, with 10 occurrences. Classification and labeling activities were a major source of concern, likely because they often serve as intermediate analytical steps that influence subsequent synthesis, comparison, and interpretation. As a result, errors introduced during these stages may propagate through the empirical workflow and affect the validity of later findings and conclusions. This concern is visible in reports where authors explicitly connected LLM-based classification to possible misclassification, unpredictability, and reduced robustness:

\begin{quote}
    \small
    \textit{``A key threat in our study is the reliance on LLMs for issue classification. Although we performed validation on a labeled subset of data, LLMs can still make misclassifications or provide unpredictable results, which could affect the robustness of our conclusions.''} [A180]
\end{quote}

The remaining lower-level themes within \textit{threats to empirical evidence construction} show that this concern extends beyond incorrect labels. \textit{LLM screening and filtering errors} and \textit{LLM-generated artifact unreliability} appeared 4 times each, while \textit{LLM extraction errors} appeared once. Together, these themes indicate that LLMs may compromise empirical evidence in different but related ways: by excluding relevant material, retaining irrelevant material, producing artifacts that inaccurately represent the original data, or extracting incomplete or imprecise information from empirical sources. In this sense, the central risk is not simply that an LLM may produce an incorrect answer, but that its output may reshape what counts as evidence in the study.

A different type of concern appears when the validity risk is tied less to the empirical artifact itself and more to the conditions under which the LLM output is produced. This is captured by \textit{threats from prompt and workflow contingency}, which accounted for 5 coded threat occurrences. This theme reflects the dependence of LLM-assisted results on prompt design, execution conditions, iterative workflows, and automation loops. Within this theme, \textit{prompt-dependent interpretation} appeared in 3 coded occurrences, showing that prompts were sometimes recognized as part of the research instrument rather than as minor implementation details. In these cases, prompt design could shape the results produced by the model, as acknowledged in the following excerpt:

\begin{quote}
    \small
    \textit{``We notice that changing the way we phrase the GPT-4 prompt gives us different results. This shows that it is not always easy to figure out the best way to phrase the prompts.''}[A926]
\end{quote}

\begin{longtblr}[
caption = {Taxonomy of LLM-related threats to validity.},
label = {tab:taxonomy-llm-threats}
]{
colspec = { Q[c,m,2cm] Q[j,m,2.5cm] Q[l,m,2cm] X[j,m] Q[l,m,2cm]
},
hlines,
row{1} = {bg=black, fg=white, font=\bfseries},
hline{1,2,5} = {0.5pt, solid},
vline{2,3,4,5} = {0.5pt, solid},
rows = {font=\small},
rowsep = 1pt
}
\textbf{Higher-level theme} &
\textbf{Description} &
\textbf{Lower-level theme} &
\textbf{Description} &
\textbf{Paper IDs} \\

\SetCell[r=4]{} Threats to empirical evidence construction &
\SetCell[r=4]{} Threats in which LLM use may compromise the construction, transformation, classification, screening, or extraction of empirical evidence used in the study. &
LLM-generated artifact unreliability &
LLM-generated or LLM-rewritten artifacts may be inaccurate, unrealistic, incomplete, or may distort the intended meaning of the original material. & A46, A671, A926, A1004\\

&
&
LLM classification and labeling errors &
LLMs may assign incorrect labels, categories, or classifications, including errors caused by hallucination, subjective interpretation, or limitations of automated classification. & A46, A64, A96, A154, A180, A447, A806, A936, A966\\

&
&
LLM screening and filtering errors &
LLMs may incorrectly include, exclude, miss, or filter relevant items during screening or selection procedures. & A125, A180, A806, A1455 \\

&
&
LLM extraction errors &
LLMs may extract incorrect, incomplete, or imprecise information from empirical sources, affecting the quality of the evidence base. & A1315 \\

Threats from LLM-based evaluation and measurement &
Threats that arise when LLMs are used to evaluate, judge, score, or measure empirical artifacts, potentially affecting the validity of the resulting assessment. &
LLM evaluation bias &
LLM-based judgments may introduce evaluator bias, misalignment with human judgment, or inaccuracies when the model is used as an assessor. & A46, A767, A779 \\

\SetCell[r=2]{} Threats from prompt and workflow contingency &
\SetCell[r=2]{} Threats caused by the dependence of LLM outputs on prompt design, iterative workflows, automation loops, or other contingent execution conditions. &
LLM workflow instability &
LLM-assisted workflows may produce unstable or unreliable outcomes when outputs depend on iterative refinement, feedback loops, or automated processing steps. & A64, A806 \\

&
&
Prompt-dependent interpretation &
LLM outputs may vary according to prompt wording, prompt templates, or prompt design choices, affecting the consistency and interpretation of results. & A776, A925, A926 \\

Threats from model contingency and domain fit &
Threats related to the dependence of results on the selected model, model version, or the degree of alignment between the model and the empirical task domain. &
Model dependence and domain mismatch &
Findings may be affected by model version instability or by a mismatch between the selected LLM and the domain, context, or task being analyzed. & A925 \\

Threats from downstream propagation of LLM outputs &
Threats in which errors or distortions introduced by LLMs in earlier stages propagate into later stages of analysis, synthesis, or reporting. &
Downstream distortion of empirical results &
LLM-generated summaries, classifications, or extractions may distort empirical findings and propagate these distortions into subsequent analyses or conclusions. & A1106 \\
\end{longtblr}

This concern becomes more pronounced when LLMs are embedded in repeated or chained workflows. The lower-level theme \textit{LLM workflow instability}, with 2 occurrences, captures risks that emerge when outputs depend on feedback loops, refinement cycles, or automated processing steps. In these cases, validity threats may arise not only from a single inaccurate response, but also from the behavior of the workflow as a whole, including error accumulation, unstable outputs across iterations, or failure of refinement loops to converge. This suggests that some LLM-related threats are workflow-level threats rather than merely output-level threats.

The taxonomy also distinguishes cases in which LLMs were used not to construct evidence, but to evaluate or measure empirical artifacts. These cases are captured by \textit{threats from LLM-based evaluation and measurement}, which accounted for 4 coded threat occurrences, all grouped under \textit{LLM evaluation bias}. This theme is methodologically important because the LLM acts as an assessor rather than merely as a support tool. When models are used to evaluate quality, judge relevance, classify correctness, or compare artifacts, their biases, inconsistencies, or misalignments with human judgment may directly influence the study's reported results. Among the studies coded under this theme, the following excerpt illustrates how authors framed LLM-based evaluation as a possible source of bias or inaccuracy:

\begin{quote}
    \small
    \textit{``Our evaluation of assertion message quality (LLM-Eval), structural classification, and test comments relied on GPT-4, potentially introducing evaluator bias or inaccuracies from automated classification.''}[A46]
\end{quote}

Other threats become visible only when considering how LLM outputs move through the broader empirical workflow. \textit{Threats from downstream propagation of LLM outputs}, with 3 coded occurrences, captures cases in which LLM-generated or LLM-mediated outputs are reused as inputs to later stages of analysis, synthesis, or reporting. The corresponding lower-level theme, \textit{downstream distortion of empirical results}, highlights that errors introduced by LLMs may not remain local to the task in which they first appear. Instead, they may propagate into derived metrics, synthesized findings, or final conclusions. Therefore, the validity impact of LLM use depends not only on whether an individual output is correct, but also on how that output is subsequently reused. This propagation risk becomes explicit when authors connect LLM-generated classifications or summaries to later quantitative or interpretive results, as illustrated by the following excerpt:

\begin{quote}
    \small
    \textit{``The use of ChatGPT in the initial classification and summarization of the scraped data introduces a potential threat to validity, particularly in the quantitative correlations presented in relation to RQ3.''}[A1106]
\end{quote}

Finally, the taxonomy includes a more structural source of validity concern: \textit{threats from model contingency and domain fit}, which accounted for 2 coded threat occurrences. These threats, grouped under \textit{model dependence and domain mismatch}, concern the dependence of results on the selected model, model version, training data, or degree of alignment between the model and the domain. This theme points to a different kind of fragility: even when a workflow is carefully designed, findings obtained with one LLM may not necessarily generalize to another model, remain stable after model updates, or transfer to tasks for which the model has limited domain fit. Model contingency was reported as a reproducibility concern when authors noted that results obtained with a specific proprietary model may change as the model is updated:

\begin{quote}
\small
\textit{``A key limitation of our classification approach involves the challenges of reproducibility with OpenAI’s proprietary GPT models, and large language models (LLMs) in general (Biderman et al. 2024; Spirling 2023). Despite the use of the most up-to-date LLM available at the time we conducted this study, updates to GPT models by OpenAI could alter our experimental outcomes, which could complicate consistent comparisons across studies (Biderman et al. 2024).''}[A925]
\end{quote}

Importantly, model contingency  relevance depends on the role played by the LLM and the extent to which the resulting evidence depends on model-specific behavior. For example, when an LLM is used only to produce labels that are independently validated against a reliable reference, variation across models or versions may have limited impact on the study findings. In contrast, model contingency becomes more consequential when LLM outputs are used without comprehensive validation, when the model participates in the operationalization or interpretation of constructs, or when its outputs are propagated into subsequent analyses. Thus, the relevant concern is not simply that different models may produce different outputs, but whether such model-dependent variation can materially alter the empirical evidence or conclusions of the study.

Viewed as a whole, the taxonomy shows that reported LLM-related threats to validity are concentrated mainly around the construction of empirical evidence, especially when LLMs classify, screen, extract, or generate the material on which later analysis depends. However, the reported threats are not limited to inaccurate outputs. They also involve prompt sensitivity, workflow instability, evaluator bias, downstream propagation of errors, and dependence on specific models or domain fit. This suggests that LLMs introduce a methodological layer that is often only partially visible in empirical reports: model choices, prompts, interaction protocols, and reuse of LLM outputs may shape the research process even when they are not treated as methodological decisions. \textbf{The central concern, therefore, is not only whether an LLM output is correct, but whether the study reports how LLM-assisted steps entered the empirical workflow, how their risks were assessed, and how their effects on the findings were controlled}.

\subsection{RQ3: Mapping LLM-related threats to established validity dimensions}

\begin{figure}[ht]
    \centering
   \includegraphics[width=\linewidth]{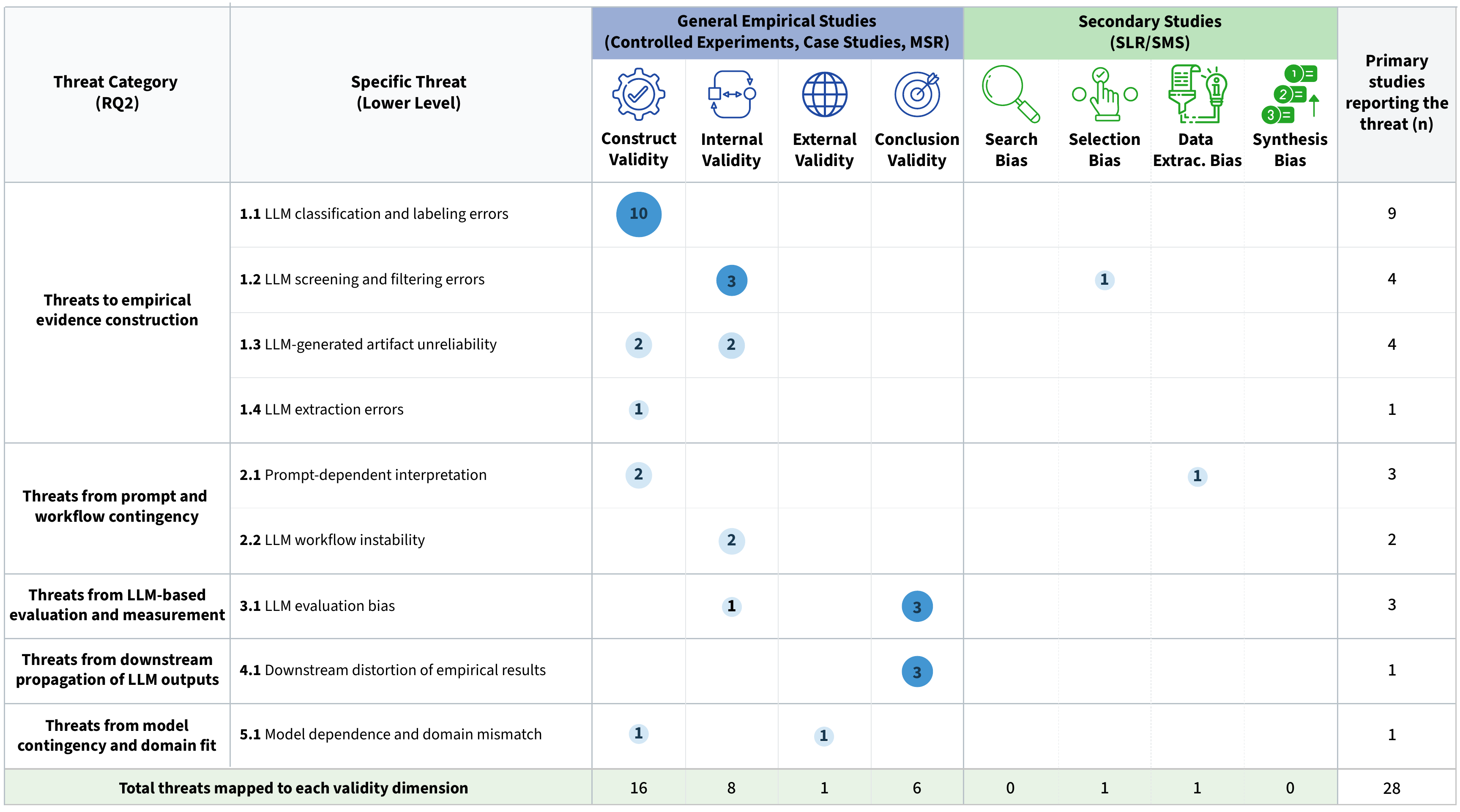}
    \caption{Mapping of reported LLM-related threat occurrences to classical validity dimensions.}
    \Description{Matrix mapping nine specific LLM-related threats, grouped into five threat categories, to validity dimensions for general empirical studies and secondary studies. General empirical studies, shown with blue headers, include construct, internal, external, and conclusion validity; secondary studies, shown with green headers, include search, selection, data extraction, and synthesis bias. Across all threats, there are 16 mappings to construct validity, 8 to internal validity, 1 to external validity, 6 to conclusion validity, 1 to selection bias, and 1 to data extraction bias; no threats are mapped to search or synthesis bias. The largest individual mapping is LLM classification and labeling errors to construct validity, with 10 instances. Other notable mappings include screening and filtering errors to internal validity (3), LLM-generated artifact unreliability to construct and internal validity (2 each), LLM evaluation bias to conclusion validity (3), and downstream distortion of empirical results to conclusion validity (3). The rightmost column reports the number of primary studies associated with each threat, totaling 28 study-level reports. Circles containing numbers indicate the frequency of mappings within each validity dimension.}
    \label{fig:task-by-threat-dimension}
\end{figure}

To answer \textbf{RQ3}, we analyzed how the threats to validity introduced by LLM-assisted research tasks identified in RQ2 manifested across the established validity dimensions associated with each study type, following the frameworks presented in Table~\ref{tab:validity-threats}. Figure~\ref{fig:task-by-threat-dimension} presents the distribution of reported threats across validity dimensions. Table~\ref{tab:llm-threats-validity-justification} further complements this analysis by providing the rationale for each mapped relationship, explaining why each LLM-related threat affects specific classical validity dimensions across primary and secondary studies.

The mapping in Figure~\ref{fig:task-by-threat-dimension} shows an uneven distribution of LLM-related threats across classical validity dimensions. Construct validity accounts for the largest share of mapped relationships, followed by internal and conclusion validity, whereas external validity and the secondary-study dimensions of search and synthesis bias are much less visible. This pattern should not be interpreted as evidence that LLM-assisted workflows mainly threaten construct validity in an absolute sense. Rather, it indicates that the threats explicitly reported in the analyzed studies are concentrated around more visible forms of representational error, such as misclassification, unreliable generated artifacts, and inaccurate extraction.

\begin{longtblr}[
    caption={Rationale for mapping LLM-related threat categories to the validity dimensions used for each study type.},
    label={tab:llm-threats-validity-justification}
] {
    colspec = {Q[j, m, 4cm] Q[c, m, 2.8cm] X[j, m]},
    hlines,
    row{1} = {bg=black, fg=white, font=\bfseries},
    rows = {font=\small},
    rowsep = 1pt
}
\textbf{Threat Category} & \textbf{Mapped Validity Dimension} & \textbf{Mapping Rationale} \\
\textbf{1.1 LLM classification and labeling errors} 
& Construct validity 
& Classification or labeling errors can assign empirical observations to categories that do not represent the intended construct, weakening the alignment between the phenomenon under study and its operationalized representation. \\

\SetCell[r=2]{} \textbf{1.2 LLM screening and filtering errors} 
& Internal validity 
& LLM-driven screening or filtering processes can introduce uncontrolled inclusion and exclusion decisions into the empirical procedure, altering the set of artifacts or data items analyzed. \\

& Selection bias 
& LLM-driven screening can systematically exclude relevant studies or retain irrelevant ones, biasing the evidence base used for review. \\

\SetCell[r=2]{} \textbf{1.3  LLM-generated artifact unreliability} 
& Construct validity 
& Generated artifacts may fail to instantiate the construct they are intended to represent, producing empirical material that only partially or inaccurately reflects the target phenomenon. \\

& Internal validity 
& Variation in the quality or reliability of generated artifacts can become an uncontrolled procedural factor that influences the results independently of the variables under study. \\

\textbf{1.4  LLM extraction errors} 
& Construct validity 
& Extraction errors can misrepresent the empirical attributes being measured or analyzed, weakening the correspondence between source material and analytical representation. \\

\SetCell[r=2]{} \textbf{2.1 Prompt-dependent interpretation} 
& Construct validity 
& Prompt wording can shift how the LLM interprets constructs, leading to inconsistent operationalization across prompts, runs, or studies. \\

& Data Extraction Bias
& When prompts guide extraction from source materials, prompt wording can influence which information is selected, omitted, or emphasized, producing incomplete or distorted extracted data. \\

\textbf{2.2 LLM workflow instability} 
& Internal validity 
& Iterative or automated workflows may introduce inconsistent intermediate decisions that are difficult to monitor or control. \\

\SetCell[r=2]{} \textbf{3.1 LLM evaluation bias} 
& Internal validity 
& When LLMs act as evaluators, their judgments can introduce systematic bias or inconsistency into the analytical procedure, affecting outcomes independently of the studied phenomenon. \\

& Conclusion validity 
& Biased LLM evaluations can distort comparisons, rankings, or assessments that support the study's claims, leading to conclusions that are not fully supported by the underlying evidence. \\

\textbf{4.1 Downstream distortion of empirical results} 
& Conclusion validity 
& Errors in LLM-generated intermediate outputs can propagate into derived metrics, interpretations, or claims, reducing the reliability of the conclusions drawn from the analysis. \\

\SetCell[r=2]{} \textbf{5.1 Model dependence and domain mismatch} 
& Construct validity 
& When the model is poorly aligned with the empirical context or domain terminology, it may interpret key concepts through model-specific approximations rather than the intended research constructs. \\
& External validity 
& Findings obtained with a specific model, version, configuration, or domain context may not transfer to other models, datasets, empirical settings, or future model releases. \\
\end{longtblr}

The strongest pattern concerns the role of LLMs in empirical evidence construction. \textit{LLM classification and labeling errors} were mapped to construct validity in 10 occurrences and reported by nine primary studies. This indicates that, when LLMs are used to assign labels, categories, or classifications, they do not merely accelerate an existing analytical step; they participate in the operationalization of the constructs under study. In these cases, a classification error is not only a local mistake in a dataset. It may alter how the phenomenon is represented, especially when LLM-generated labels are later used as variables, categories, evidence units, or inputs to subsequent analyses. The concentration under construct validity therefore points to a broader methodological concern: LLMs can become implicit instruments of measurement, even when they are described as tools for automation or assistance.

The same concern helps explain why \textit{prompt-dependent interpretation}, \textit{LLM extraction errors}, and \textit{model dependence and domain mismatch} were also mapped to construct validity. In traditional empirical workflows, construct validity is usually threatened by problems in instruments, coding schemes, measurement procedures, or operational definitions. In LLM-assisted workflows, these elements may be partially embedded in prompts, model behavior, and interaction protocols. As a result, the operational meaning of a construct may be shaped not only by researchers' explicit design decisions, but also by how the model interprets instructions, resolves ambiguity, and maps textual input to output categories.

Internal and conclusion validity reveal a different aspect of the problem. While construct-validity mappings concern whether empirical objects are represented correctly, internal-validity mappings point to procedural instability inside the research workflow. Threats such as \textit{LLM screening and filtering errors}, \textit{LLM-generated artifact unreliability}, \textit{LLM workflow instability}, and \textit{LLM evaluation bias} show that LLMs may introduce uncontrolled variation into the empirical procedure. This variation may remain hidden behind apparently standardized workflows, since two studies can report the same type of LLM-assisted task while differing in prompt wording, model version, configuration settings, sampling behavior, interaction sequence, or iterative refinement strategy.

Conclusion-validity mappings show how these risks can propagate into final claims. \textit{LLM evaluation bias} and \textit{downstream distortion of empirical results} illustrate that LLM-mediated decisions do not necessarily remain confined to the stage in which they first appear. A biased judgment, inaccurate classification, unreliable generated artifact, or distorted intermediate summary may later influence rankings, comparisons, derived metrics, or reported interpretations. For this reason, LLM-related threats are not always well captured as isolated validity problems. They may operate as chains of methodological dependency, where a problem introduced during representation, extraction, generation, or evaluation later affects the credibility of the study's conclusions.

External validity was rarely reported, and only \textit{model dependence and domain mismatch} was mapped to this dimension. This low frequency is notable because many LLM-assisted studies rely on specific models, versions, prompts, and configurations whose behavior may not generalize across tools, domains, datasets, or future model releases. The limited reporting of external-validity threats may therefore reflect a blind spot in the current literature rather than a lower level of methodological risk. Existing studies appear to focus more on whether an LLM-assisted procedure works in a local setting than on whether the findings would remain stable under different models, model versions, domains, or execution conditions.

The secondary-study dimensions show a sharper reporting gap. Reported threats were mapped only to \textit{selection bias} and \textit{data extraction bias}, while no explicit threats were mapped to \textit{search bias} or \textit{synthesis bias}. This does not imply that LLM use cannot affect search or synthesis in secondary studies. Rather, within the analyzed corpus, explicit LLM-related threats were attached only to tasks connected to study screening, study filtering, and data extraction, even though the secondary-study tasks also included activities closer to study identification, topic identification, relevance assessment, and evidence retrieval. This suggests that reporting is more explicit around discrete review steps where LLM outputs can be inspected against inclusion criteria or source documents, whereas threats associated with how LLMs may shape the discovery, prioritization, or organization of evidence remain less documented.

A related reporting gap appears in survey studies. Although three survey studies in the corpus employed LLM-assisted tasks, none explicitly reported LLM-related threats to validity. This absence is relevant because the survey studies used LLMs for tasks such as persona-based generation of survey responses, thematic analysis of textual survey results, and rewriting research artifacts according to a predefined template. The first two uses are especially important because they can influence how participant data are generated, represented, interpreted, or transformed into evidence. The lack of reported threats therefore suggests that survey-based empirical software engineering may not yet have a well-developed vocabulary for identifying how LLM assistance can affect response validity, qualitative interpretation, and the transformation of survey-related materials into empirical findings.

Overall, the mapping shows that \textbf{classical validity dimensions provide a useful structure for locating LLM-related threats, but it also exposes the limits of treating those dimensions as isolated categories}. 

\section{Discussion}
\label{sec:discussion}

The results of this study provide complementary perspectives on how the hidden methodological layer manifests in LLM-assisted empirical software engineering. While the previous section examined the visibility, nature, and validity impact of reported threats to validity introduced by LLM-assisted research tasks, the findings collectively suggest a broader methodological shift introduced by the integration of LLMs into empirical workflows. In this section, we synthesize these results through the lens of the proposed conceptual model, discussing how LLM-assisted workflows impact traditional assumptions regarding methodological transparency, controllability, reproducibility, and validity in empirical software engineering research. We further discuss the cross-dimensional nature of threats to validity introduced by LLM-assisted research tasks and the implications of these findings for future empirical software engineering practices and methodological standards.

\subsection{The Hidden Methodological Layer}

The findings of this study suggest that LLM-assisted ESE cannot be fully understood by looking only at the visible sequence of empirical activities reported in a study. When LLMs are used to classify artifacts, screen studies, extract information, generate or transform empirical material, evaluate outputs, or synthesize findings, part of the empirical procedure is delegated to a model-mediated process. In these cases, the LLM is not merely an external productivity tool; it becomes part of the methodological apparatus through which evidence is constructed, interpreted, and reused.

We refer to this phenomenon as the \textbf{hidden methodological layer}. This layer captures the model- and prompt-mediated conditions through which LLMs influence empirical work, including prompt design, model configuration, interaction history, context construction, iterative refinement, automation decisions, validation procedures, and dependencies on specific model versions or capabilities. The layer is ``hidden'' not because these elements are necessarily inaccessible, but because their methodological influence is often indirect, only partially inspectable, or insufficiently reported as part of the research method. A prompt, for example, may encode assumptions about a construct; a model may interpret domain-specific terminology in ways not anticipated by researchers; and an iterative interaction may progressively reshape how empirical evidence is represented. Thus, methodological decisions are no longer located only in explicit protocols, instruments, or coding schemes, but are distributed across the interaction between researchers, models, and generated outputs.

\begin{figure}[ht]
    \centering
    \includegraphics[width=\linewidth]{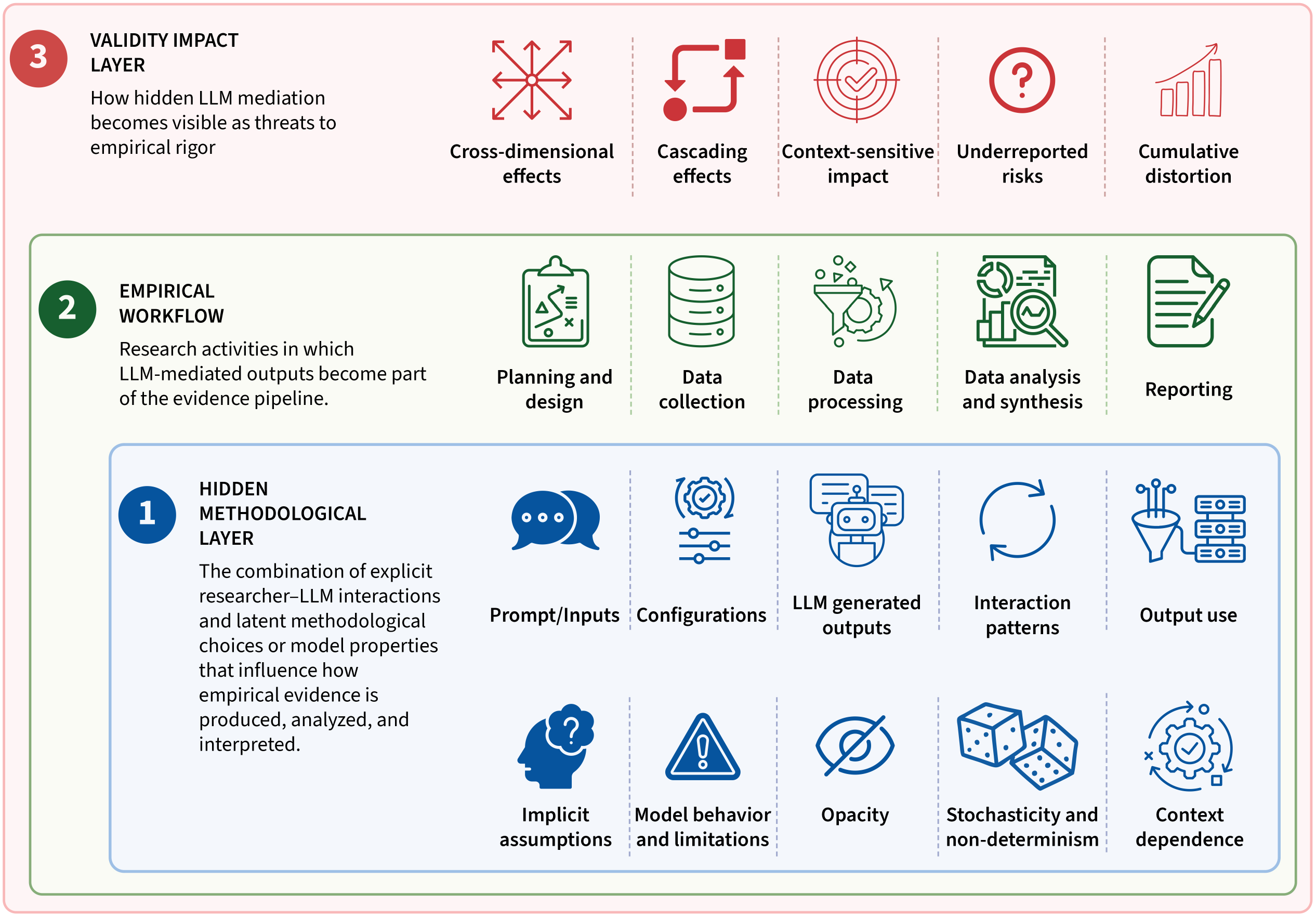}
    \caption{Conceptual model of the hidden methodological layer in LLM-assisted empirical workflows.}
    \Description{Three-layer conceptual framework illustrating how LLM use can affect empirical research. The blue Hidden Methodological Layer contains researcher–LLM interactions and latent properties such as prompts, configurations, generated outputs, opacity, stochasticity, and context dependence. These can influence the green Empirical Workflow, spanning planning, data collection, processing, analysis, and reporting. Their effects may emerge in the red Validity Impact Layer as cross-dimensional, cascading, context-sensitive, underreported, or cumulative threats to empirical rigor.}
    \label{fig:hiddenMethodologicalLayerDiagram}
\end{figure}

Figure~\ref{fig:hiddenMethodologicalLayerDiagram} illustrates this layered structure. Rather than representing LLM-assisted empirical research as a linear sequence of tasks, the model shows how visible empirical workflows are supported by deeper interactional and methodological conditions. The hidden methodological layer makes the model operational through prompts, configurations, outputs, iterations, and reuse of generated artifacts. Moreover, it also captures the less visible conditions that shape model behavior, such as implicit assumptions, opacity, stochasticity, context dependence, and model limitations. The outer layer represents the validity impact: LLM mediation may become visible through cross-dimensional, cascading, context-sensitive, underreported, or cumulative effects on validity.

This model helps clarify why LLM-related validity threats are difficult to localize. In conventional empirical workflows, threats to validity are often associated with identifiable methodological components, such as sampling decisions, measurement instruments, coding procedures, or analysis techniques. In LLM-assisted workflows, however, validity-relevant decisions may be embedded in prompts, model responses, interaction sequences, or downstream uses of generated outputs. As a result, a threat may not appear as a single visible methodological flaw. It may emerge from the interaction between the task assigned to the LLM, the model's behavior, the researchers' interpretation of its output, and the way that output is incorporated into later stages of the study.

This shift also changes how output errors should be interpreted. The methodological concern is not limited to whether a specific LLM response is correct or incorrect. An incorrect label, summary, extraction, or evaluation becomes more consequential when it enters the evidence pipeline and is reused as a basis for further analysis. In such cases, the LLM output is not simply an isolated artifact; it becomes part of the empirical infrastructure of the study. This helps explain why several threats identified in our taxonomy involve not only visible output errors, but also prompt-dependent interpretation, workflow instability, evaluator bias, model dependence, and downstream distortion of empirical results.

The hidden methodological layer also helps interpret the underreporting observed in RQ1. If LLM-mediated decisions are treated as implementation details rather than methodological decisions, their validity implications may remain absent from the threats-to-validity discussion. This absence is problematic because it limits readers' ability to assess how LLM outputs were produced, whether they were validated, how sensitive they were to prompts or model configurations, and how they affected subsequent analyses. In this sense, underreporting is not merely a documentation problem; it obscures part of the empirical process through which evidence is generated and transformed.

At the same time, the hidden methodological layer should not be interpreted as evidence that LLM-assisted workflows are inherently invalid or unreliable. Rather, it indicates that traditional assumptions about transparency, determinism, and controllability become less stable when LLMs participate in empirical tasks. Established validity concepts remain applicable, but the mechanisms through which validity threats arise may become more dynamic, interaction-dependent, and difficult to inspect. This is why LLM-assisted workflows require reporting and validation practices that make model-mediated methodological conditions visible enough to be assessed.

The central implication is that LLM use should be reported as part of the research method, not only as a tool choice. Making the hidden methodological layer explicit requires documenting prompts, model configurations, interaction protocols and validation procedures. More importantly, it requires treating these elements as validity-relevant methodological decisions. Doing so can help researchers identify hidden assumptions, evaluate robustness, trace methodological dependencies, and develop reporting standards appropriate for LLM-assisted empirical software engineering.

\subsection{Validity Implications of the Hidden Methodological Layer}

The hidden methodological layer reconfigures how validity threats emerge, become visible, and propagate in LLM-assisted empirical studies. Rather than replacing established validity concerns, LLM use changes the mechanisms through which those concerns materialize. A construct-validity concern, for example, may arise not only from a poorly defined coding scheme, but also from the way a prompt leads the model to interpret a category. An internal-validity concern may arise not only from uncontrolled study conditions, but also from variation introduced by model configuration, stochastic generation, or iterative prompting. A conclusion-validity concern may arise when LLM-generated classifications, summaries, or evaluations are reused as inputs to later analyses and eventually shape the claims drawn from the study.

This shift means that validity assessment cannot focus only on observable workflow steps or final outputs. Researchers and reviewers also need to examine the LLM interaction conditions through which outputs are produced: how prompts were designed, how models were selected and configured, how outputs were validated, how many iterations were performed, and how generated artifacts were reused downstream. In this sense, prompts can function as methodological instruments, model settings as methodological variables, and interaction patterns as part of the empirical procedure.

A central implication of this view is that LLM-related threats may be cross-dimensional. A single threat can affect different validity dimensions depending on where it occurs in the workflow and how its outputs are used. For example, prompt-dependent interpretation may affect construct validity when it changes how a concept is operationalized, but it may also affect internal validity when different prompts produce inconsistent analytical behavior. Similarly, an unreliable classification or extraction may begin as a representational problem but later affect conclusion validity if it is used to produce metrics, syntheses, comparisons, or interpretations.

LLM-related threats may also be cascading. Errors or distortions introduced early in the workflow can accumulate as LLM-generated artifacts are reused, refined, summarized, or incorporated into later stages of analysis. This is particularly important because small intermediate errors may appear harmless when inspected locally, but become more consequential when they shape subsequent analytical decisions. The validity impact of LLM-assisted workflows therefore depends not only on the quality of individual outputs, but also on the role those outputs play within the broader evidence pipeline.

These characteristics help explain the reporting gaps observed in this study. Threats that appear as visible output-level errors, such as misclassification, inaccurate extraction, or unreliable generation, are easier to identify and report. Threats embedded in interaction histories, prompt assumptions, model dependence, configuration choices, or downstream reuse are less visible and may therefore remain underreported. This suggests that reporting practices need to move beyond describing that an LLM was used and toward documenting how LLM-mediated decisions entered the empirical workflow.

\subsection{Implications for Empirical Software Engineering Research}

The findings of this study have implications for how LLM-assisted empirical software engineering should be designed, reported, evaluated, and reviewed. As LLMs become increasingly integrated into empirical workflows, methodological rigor cannot rely solely on assumptions inherited from human-driven or deterministic procedures. Instead, researchers need to explicitly account for the hidden methodological mechanisms introduced through LLM interactions and their potential impact on empirical validity.

A first implication concerns the \textbf{reporting of LLM-assisted empirical workflows}. The results of RQ1 showed that many studies employing LLMs did not explicitly report threats to validity associated with their use, even in workflows involving substantial automation or evaluation activities. This suggests that existing reporting practices may not yet adequately capture the methodological implications of LLM integration. \textbf{To improve transparency and reproducibility, future studies need to document not only the empirical procedures and tasks performed with LLM assistance, but also the model- and interaction-specific conditions under which these tasks were carried out.} Such information may include prompt engineering strategies, model versions, parameter settings, interaction protocols, validation procedures, human oversight mechanisms, and the extent to which LLM-generated outputs were reused in downstream analyses.

A second implication concerns \textbf{reproducibility and methodological stability}. Traditional empirical workflows often assume that analytical procedures can be reproduced from explicit methodological descriptions. However, the probabilistic and prompt-sensitive nature of LLMs challenges this assumption. Identical prompts may produce different outputs across executions, while small variations in prompt wording, model configuration, or interaction history may substantially alter empirical outcomes. As a result, reproducibility in LLM-assisted empirical software engineering may require additional methodological controls, such as prompt versioning, repeated executions, sensitivity analyses, configuration tracking, and explicit documentation of model dependencies.

The study also has implications for \textbf{how empirical results involving LLMs should be evaluated and interpreted}. Our taxonomy of threats to validity introduced by LLM-assisted research tasks revealed that methodological risks frequently emerge during the construction, transformation, classification, extraction, and evaluation of empirical evidence. Consequently, researchers and reviewers may need to critically examine not only the final outputs generated by LLM-assisted workflows, but also the processes through which these outputs were produced and validated. In particular, studies using LLMs for classification, synthesis, or evaluation activities may benefit from systematic human validation, cross-model comparisons, consistency checks, and robustness analyses designed to assess the stability and reliability of generated outputs.

Another implication concerns \textbf{methodological literacy}. Researchers using LLMs in empirical software engineering need to understand not only the empirical method they are applying, but also the operational properties of the models they incorporate into that method. This includes recognizing prompts as methodological instruments, model configurations as sources of variation, stochastic outputs as potential threats to stability, and generated artifacts as evidence that may require validation before reuse.

Finally, the findings suggest the need for \textbf{methodological guidelines and evaluation standards} tailored to LLM-assisted empirical research like the recent guidelines proposed by Baltes et al.~\cite{baltes2026guidelinesempiricalstudiessoftware}. Existing validity frameworks remain highly relevant, but they provide limited guidance on how to document and control risks introduced by probabilistic, opaque, and rapidly evolving AI systems. Future guidance should therefore address prompt reporting, model transparency, workflow traceability, output validation, robustness evaluation, and the management of cascading validity effects across empirical workflows.

The broader implication is that LLM integration in empirical software engineering is not only a tooling evolution. It is also a methodological challenge. Ensuring empirical rigor in LLM-assisted workflows depends on making hidden methodological mechanisms visible enough to be inspected, reported, validated, and controlled throughout the research process.

\section{Threats to Validity}
\label{sec:threats}

As this study builds on a previously conducted SLR \cite{gomes2026llmassistedESE}, its validity is influenced by both the original review protocol and the additional data collection and analysis performed in this work. To mitigate potential threats, we relied on an established protocol for study selection and ensured consistency by reusing the same corpus of primary studies.

However, the present study introduces additional analysis procedures and interpretive abstractions that create new validity considerations beyond those discussed in the original SLR. In particular, this work synthesizes and interprets threats to validity introduced by LLM-assisted research tasks, analyzes their relationships with classical validity dimensions, and proposes the concept of a hidden methodological layer. These activities required additional qualitative interpretation and conceptual abstraction, which may introduce researcher bias in the categorization, naming, and mapping of threats.

To mitigate these risks, the analysis was conducted iteratively through collaborative discussions among the authors, including repeated refinement of threat definitions, mappings, and conceptual relationships. The proposed taxonomy and mappings were grounded in explicit evidence extracted from the primary studies and continuously revisited to preserve consistency between the empirical observations and the resulting abstractions. Nevertheless, alternative interpretations of some threats, relationships, or categorizations may still be possible, particularly because many primary studies report LLM-related limitations with varying levels of methodological detail.

Another threat concerns the interpretation of cross-dimensional relationships among validity threats. While this work argues that LLM-related threats may propagate across multiple validity dimensions, these relationships were derived through analytical synthesis rather than controlled empirical validation. Consequently, the proposed propagation and interaction effects should be interpreted as theoretically grounded conceptual relationships supported by the analyzed evidence, rather than empirically demonstrated causal mechanisms.

Finally, the study is limited by the current maturity of the literature on LLM-assisted empirical software engineering. Many analyzed studies provide incomplete reporting regarding prompts, configurations, validation procedures, or interaction protocols, which constrains the visibility of the hidden methodological layer itself. As a result, some LLM-related threats may remain underreported or insufficiently documented in the analyzed corpus.

\section{Conclusion}
\label{sec:conclusion}

The increasing adoption of LLMs in empirical software engineering introduces methodological challenges that extend beyond those traditionally associated with deterministic tools and human-driven workflows. In this paper, we argued that LLM-assisted empirical workflows introduce a hidden methodological layer in which important methodological influences become partially embedded within prompts, model configurations, interaction patterns, and probabilistic model behaviors.

To investigate this phenomenon, we analyzed 50 primary studies involving 69 LLM-assisted empirical research tasks. Our results showed that LLM-related validity threats remain substantially underreported in the current literature, despite the growing use of LLMs in empirical workflows. We also developed a taxonomy of recurring threats to validity introduced by LLM-assisted research tasks, including classification and labeling errors, prompt-dependent interpretation, workflow instability, evaluator bias, downstream propagation of outputs, and model dependence. Furthermore, our analysis revealed that these threats span multiple validity dimensions, with some threats propagating across different stages of the empirical workflow.

Building on these findings, we introduced the concept of the hidden methodological layer as a conceptual lens for understanding how LLMs influence empirical software engineering research. Rather than replacing traditional validity concerns, this perspective highlights how LLM-assisted workflows reshape the mechanisms through which validity threats emerge, propagate, and interact throughout empirical studies. As LLMs and AI take on not just more research tasks but a broader variety of them, the hidden methodological layers they create will expand in both size and complexity.

More broadly, our findings suggest that the increasing integration of LLMs into empirical software engineering may require the development of LLM-aware methodological practices capable of addressing probabilistic behavior, hidden interaction mechanisms, workflow-level propagation effects, and evolving model ecosystems. A direction for future work is to empirically investigate the extent to which the identified threats affect study outcomes. Replication studies of existing LLM-assisted empirical research could systematically address or mitigate these threats and compare the resulting findings with those originally reported. Such studies could help determine whether stronger methodological controls lead not only to more robust and reproducible results, but also to different or more insightful empirical findings. Understanding when and how these hidden methodological influences materially affect empirical findings represents an important next step toward more rigorous, transparent, and trustworthy LLM-assisted empirical software engineering.

\bibliographystyle{ACM-Reference-Format}
\bibliography{refs}


\begin{thebibliography}{33}


\ifx \showCODEN    \undefined \def \showCODEN     #1{\unskip}     \fi
\ifx \showISBNx    \undefined \def \showISBNx     #1{\unskip}     \fi
\ifx \showISBNxiii \undefined \def \showISBNxiii  #1{\unskip}     \fi
\ifx \showISSN     \undefined \def \showISSN      #1{\unskip}     \fi
\ifx \showLCCN     \undefined \def \showLCCN      #1{\unskip}     \fi
\ifx \shownote     \undefined \def \shownote      #1{#1}          \fi
\ifx \showarticletitle \undefined \def \showarticletitle #1{#1}   \fi
\ifx \showURL      \undefined \def \showURL       {\relax}        \fi
\providecommand\bibfield[2]{#2}
\providecommand\bibinfo[2]{#2}
\providecommand\natexlab[1]{#1}
\providecommand\showeprint[2][]{arXiv:#2}

\bibitem[Baltes et~al\mbox{.}(2026)]%
        {baltes2026guidelinesempiricalstudiessoftware}
\bibfield{author}{\bibinfo{person}{Sebastian Baltes}, \bibinfo{person}{Florian Angermeir}, \bibinfo{person}{Chetan Arora}, \bibinfo{person}{Marvin~Muñoz Barón}, \bibinfo{person}{Chunyang Chen}, \bibinfo{person}{Lukas Böhme}, \bibinfo{person}{Fabio Calefato}, \bibinfo{person}{Neil Ernst}, \bibinfo{person}{Davide Falessi}, \bibinfo{person}{Brian Fitzgerald}, \bibinfo{person}{Davide Fucci}, \bibinfo{person}{Junda He}, \bibinfo{person}{Christoph Treude}, \bibinfo{person}{Marcos Kalinowski}, \bibinfo{person}{Stefano Lambiase}, \bibinfo{person}{Daniel Russo}, \bibinfo{person}{Mircea Lungu}, \bibinfo{person}{Cristina~Martinez Montes}, \bibinfo{person}{Lutz Prechelt}, \bibinfo{person}{Paul Ralph}, \bibinfo{person}{Rijnard van Tonder}, {and} \bibinfo{person}{Stefan Wagner}.} \bibinfo{year}{2026}\natexlab{}.
\newblock \bibinfo{title}{Guidelines for Empirical Studies in Software Engineering involving Large Language Models}.
\newblock
\showeprint[arxiv]{2508.15503}~[cs.SE]
\urldef\tempurl%
\url{https://arxiv.org/abs/2508.15503}
\showURL{%
\tempurl}


\bibitem[Basili et~al\mbox{.}(1999)]%
        {Basili1999}
\bibfield{author}{\bibinfo{person}{V.R. Basili}, \bibinfo{person}{F. Shull}, {and} \bibinfo{person}{F. Lanubile}.} \bibinfo{year}{1999}\natexlab{}.
\newblock \showarticletitle{Building knowledge through families of experiments}.
\newblock \bibinfo{journal}{\emph{IEEE Transactions on Software Engineering}} \bibinfo{volume}{25}, \bibinfo{number}{4} (\bibinfo{year}{1999}), \bibinfo{pages}{456--473}.
\newblock
\href{https://doi.org/10.1109/32.799939}{doi:\nolinkurl{10.1109/32.799939}}


\bibitem[Bender et~al\mbox{.}(2021)]%
        {bender2021stochastic}
\bibfield{author}{\bibinfo{person}{Emily~M. Bender}, \bibinfo{person}{Timnit Gebru}, \bibinfo{person}{Angelina McMillan-Major}, {and} \bibinfo{person}{Shmargaret Shmitchell}.} \bibinfo{year}{2021}\natexlab{}.
\newblock \showarticletitle{On the Dangers of Stochastic Parrots: Can Language Models Be Too Big?}. In \bibinfo{booktitle}{\emph{Proceedings of the 2021 ACM Conference on Fairness, Accountability, and Transparency}} (Virtual Event, Canada) \emph{(\bibinfo{series}{FAccT '21})}. \bibinfo{publisher}{Association for Computing Machinery}, \bibinfo{address}{New York, NY, USA}, \bibinfo{pages}{610–623}.
\newblock
\showISBNx{9781450383097}
\href{https://doi.org/10.1145/3442188.3445922}{doi:\nolinkurl{10.1145/3442188.3445922}}


\bibitem[Bommasani et~al\mbox{.}(2022)]%
        {Bommasani2021}
\bibfield{author}{\bibinfo{person}{Rishi Bommasani}, \bibinfo{person}{Drew~A. Hudson}, \bibinfo{person}{Ehsan Adeli}, \bibinfo{person}{Russ Altman}, \bibinfo{person}{Simran Arora}, \bibinfo{person}{Sydney von Arx}, \bibinfo{person}{Michael~S. Bernstein}, \bibinfo{person}{Jeannette Bohg}, \bibinfo{person}{Antoine Bosselut}, \bibinfo{person}{Emma Brunskill}, \bibinfo{person}{Erik Brynjolfsson}, \bibinfo{person}{Shyamal Buch}, \bibinfo{person}{Dallas Card}, \bibinfo{person}{Rodrigo Castellon}, \bibinfo{person}{Niladri Chatterji}, \bibinfo{person}{Annie Chen}, \bibinfo{person}{Kathleen Creel}, \bibinfo{person}{Jared~Quincy Davis}, \bibinfo{person}{Dora Demszky}, \bibinfo{person}{Chris Donahue}, \bibinfo{person}{Moussa Doumbouya}, \bibinfo{person}{Esin Durmus}, \bibinfo{person}{Stefano Ermon}, \bibinfo{person}{John Etchemendy}, \bibinfo{person}{Kawin Ethayarajh}, \bibinfo{person}{Li Fei-Fei}, \bibinfo{person}{Chelsea Finn}, \bibinfo{person}{Trevor Gale}, \bibinfo{person}{Lauren Gillespie}, \bibinfo{person}{Karan
  Goel}, \bibinfo{person}{Noah Goodman}, \bibinfo{person}{Shelby Grossman}, \bibinfo{person}{Neel Guha}, \bibinfo{person}{Tatsunori Hashimoto}, \bibinfo{person}{Peter Henderson}, \bibinfo{person}{John Hewitt}, \bibinfo{person}{Daniel~E. Ho}, \bibinfo{person}{Jenny Hong}, \bibinfo{person}{Kyle Hsu}, \bibinfo{person}{Jing Huang}, \bibinfo{person}{Thomas Icard}, \bibinfo{person}{Saahil Jain}, \bibinfo{person}{Dan Jurafsky}, \bibinfo{person}{Pratyusha Kalluri}, \bibinfo{person}{Siddharth Karamcheti}, \bibinfo{person}{Geoff Keeling}, \bibinfo{person}{Fereshte Khani}, \bibinfo{person}{Omar Khattab}, \bibinfo{person}{Pang~Wei Koh}, \bibinfo{person}{Mark Krass}, \bibinfo{person}{Ranjay Krishna}, \bibinfo{person}{Rohith Kuditipudi}, \bibinfo{person}{Ananya Kumar}, \bibinfo{person}{Faisal Ladhak}, \bibinfo{person}{Mina Lee}, \bibinfo{person}{Tony Lee}, \bibinfo{person}{Jure Leskovec}, \bibinfo{person}{Isabelle Levent}, \bibinfo{person}{Xiang~Lisa Li}, \bibinfo{person}{Xuechen Li}, \bibinfo{person}{Tengyu Ma},
  \bibinfo{person}{Ali Malik}, \bibinfo{person}{Christopher~D. Manning}, \bibinfo{person}{Suvir Mirchandani}, \bibinfo{person}{Eric Mitchell}, \bibinfo{person}{Zanele Munyikwa}, \bibinfo{person}{Suraj Nair}, \bibinfo{person}{Avanika Narayan}, \bibinfo{person}{Deepak Narayanan}, \bibinfo{person}{Ben Newman}, \bibinfo{person}{Allen Nie}, \bibinfo{person}{Juan~Carlos Niebles}, \bibinfo{person}{Hamed Nilforoshan}, \bibinfo{person}{Julian Nyarko}, \bibinfo{person}{Giray Ogut}, \bibinfo{person}{Laurel Orr}, \bibinfo{person}{Isabel Papadimitriou}, \bibinfo{person}{Joon~Sung Park}, \bibinfo{person}{Chris Piech}, \bibinfo{person}{Eva Portelance}, \bibinfo{person}{Christopher Potts}, \bibinfo{person}{Aditi Raghunathan}, \bibinfo{person}{Rob Reich}, \bibinfo{person}{Hongyu Ren}, \bibinfo{person}{Frieda Rong}, \bibinfo{person}{Yusuf Roohani}, \bibinfo{person}{Camilo Ruiz}, \bibinfo{person}{Jack Ryan}, \bibinfo{person}{Christopher Ré}, \bibinfo{person}{Dorsa Sadigh}, \bibinfo{person}{Shiori Sagawa},
  \bibinfo{person}{Keshav Santhanam}, \bibinfo{person}{Andy Shih}, \bibinfo{person}{Krishnan Srinivasan}, \bibinfo{person}{Alex Tamkin}, \bibinfo{person}{Rohan Taori}, \bibinfo{person}{Armin~W. Thomas}, \bibinfo{person}{Florian Tramèr}, \bibinfo{person}{Rose~E. Wang}, \bibinfo{person}{William Wang}, \bibinfo{person}{Bohan Wu}, \bibinfo{person}{Jiajun Wu}, \bibinfo{person}{Yuhuai Wu}, \bibinfo{person}{Sang~Michael Xie}, \bibinfo{person}{Michihiro Yasunaga}, \bibinfo{person}{Jiaxuan You}, \bibinfo{person}{Matei Zaharia}, \bibinfo{person}{Michael Zhang}, \bibinfo{person}{Tianyi Zhang}, \bibinfo{person}{Xikun Zhang}, \bibinfo{person}{Yuhui Zhang}, \bibinfo{person}{Lucia Zheng}, \bibinfo{person}{Kaitlyn Zhou}, {and} \bibinfo{person}{Percy Liang}.} \bibinfo{year}{2022}\natexlab{}.
\newblock \bibinfo{title}{On the Opportunities and Risks of Foundation Models}.
\newblock
\showeprint[arxiv]{2108.07258}~[cs.LG]
\urldef\tempurl%
\url{https://arxiv.org/abs/2108.07258}
\showURL{%
\tempurl}


\bibitem[Brown et~al\mbox{.}(2020)]%
        {Brown2020language}
\bibfield{author}{\bibinfo{person}{Tom~B. Brown}, \bibinfo{person}{Benjamin Mann}, \bibinfo{person}{Nick Ryder}, \bibinfo{person}{Melanie Subbiah}, \bibinfo{person}{Jared Kaplan}, \bibinfo{person}{Prafulla Dhariwal}, \bibinfo{person}{Arvind Neelakantan}, \bibinfo{person}{Pranav Shyam}, \bibinfo{person}{Girish Sastry}, \bibinfo{person}{Amanda Askell}, \bibinfo{person}{Sandhini Agarwal}, \bibinfo{person}{Ariel Herbert-Voss}, \bibinfo{person}{Gretchen Krueger}, \bibinfo{person}{Tom Henighan}, \bibinfo{person}{Rewon Child}, \bibinfo{person}{Aditya Ramesh}, \bibinfo{person}{Daniel~M. Ziegler}, \bibinfo{person}{Jeffrey Wu}, \bibinfo{person}{Clemens Winter}, \bibinfo{person}{Christopher Hesse}, \bibinfo{person}{Mark Chen}, \bibinfo{person}{Eric Sigler}, \bibinfo{person}{Mateusz Litwin}, \bibinfo{person}{Scott Gray}, \bibinfo{person}{Benjamin Chess}, \bibinfo{person}{Jack Clark}, \bibinfo{person}{Christopher Berner}, \bibinfo{person}{Sam McCandlish}, \bibinfo{person}{Alec Radford}, \bibinfo{person}{Ilya Sutskever},
  {and} \bibinfo{person}{Dario Amodei}.} \bibinfo{year}{2020}\natexlab{}.
\newblock \showarticletitle{Language models are few-shot learners}. In \bibinfo{booktitle}{\emph{Proceedings of the 34th International Conference on Neural Information Processing Systems}} (Vancouver, BC, Canada) \emph{(\bibinfo{series}{NIPS '20})}. \bibinfo{publisher}{Curran Associates Inc.}, \bibinfo{address}{Red Hook, NY, USA}, Article \bibinfo{articleno}{159}, \bibinfo{numpages}{25}~pages.
\newblock
\showISBNx{9781713829546}


\bibitem[Cruzes and Dyba(2011)]%
        {ThematicAnalysisSE}
\bibfield{author}{\bibinfo{person}{Daniela~S. Cruzes} {and} \bibinfo{person}{Tore Dyba}.} \bibinfo{year}{2011}\natexlab{}.
\newblock \showarticletitle{Recommended Steps for Thematic Synthesis in Software Engineering}. In \bibinfo{booktitle}{\emph{Proceedings of the 2011 International Symposium on Empirical Software Engineering and Measurement}} \emph{(\bibinfo{series}{ESEM '11})}. \bibinfo{publisher}{IEEE Computer Society}, \bibinfo{address}{USA}, \bibinfo{pages}{275–284}.
\newblock
\showISBNx{9780769546049}
\href{https://doi.org/10.1109/ESEM.2011.36}{doi:\nolinkurl{10.1109/ESEM.2011.36}}


\bibitem[De~Morais~Le\c{c}a et~al\mbox{.}(2025)]%
        {demoraisleca2025apps}
\bibfield{author}{\bibinfo{person}{Matheus De~Morais~Le\c{c}a}, \bibinfo{person}{Lucas Valen\c{c}a}, \bibinfo{person}{Reydne Santos}, {and} \bibinfo{person}{Ronnie De~Souza~Santos}.} \bibinfo{year}{2025}\natexlab{}.
\newblock \showarticletitle{Applications and Implications of Large Language Models in Qualitative Analysis: A New Frontier for Empirical Software Engineering}. In \bibinfo{booktitle}{\emph{2025 IEEE/ACM International Workshop on Methodological Issues with Empirical Studies in Software Engineering (WSESE)}} (Ottawa, ON, Canada). \bibinfo{publisher}{IEEE Press}, \bibinfo{pages}{36–43}.
\newblock
\href{https://doi.org/10.1109/WSESE66602.2025.00013}{doi:\nolinkurl{10.1109/WSESE66602.2025.00013}}


\bibitem[Easterbrook et~al\mbox{.}(2008)]%
        {Easterbrook2008}
\bibfield{author}{\bibinfo{person}{Steve Easterbrook}, \bibinfo{person}{Janice Singer}, \bibinfo{person}{Margaret-Anne Storey}, {and} \bibinfo{person}{Daniela Damian}.} \bibinfo{year}{2008}\natexlab{}.
\newblock \showarticletitle{Selecting Empirical Methods for Software Engineering Research}.
\newblock In \bibinfo{booktitle}{\emph{Guide to Advanced Empirical Software Engineering}}, \bibfield{editor}{\bibinfo{person}{Forrest Shull}, \bibinfo{person}{Janice Singer}, {and} \bibinfo{person}{Dag I.~K. Sj{\o}berg}} (Eds.). \bibinfo{publisher}{Springer London}, \bibinfo{address}{London}, \bibinfo{pages}{285--311}.
\newblock
\showISBNx{978-1-84800-044-5}
\href{https://doi.org/10.1007/978-1-84800-044-5_11}{doi:\nolinkurl{10.1007/978-1-84800-044-5_11}}


\bibitem[Gilardi et~al\mbox{.}(2023)]%
        {Gilardi2023}
\bibfield{author}{\bibinfo{person}{Fabrizio Gilardi}, \bibinfo{person}{Meysam Alizadeh}, {and} \bibinfo{person}{Ma{\"e}l Kubli}.} \bibinfo{year}{2023}\natexlab{}.
\newblock \showarticletitle{{ChatGPT} Outperforms Crowd-Workers for Text-Annotation Tasks}.
\newblock \bibinfo{journal}{\emph{Proceedings of the National Academy of Sciences}} \bibinfo{volume}{120}, \bibinfo{number}{30} (\bibinfo{year}{2023}).
\newblock
\href{https://doi.org/10.1073/pnas.2305016120}{doi:\nolinkurl{10.1073/pnas.2305016120}}


\bibitem[Gomes et~al\mbox{.}(2026a)]%
        {replicationPackage}
\bibfield{author}{\bibinfo{person}{Vict{\'o}ria Gomes}, \bibinfo{person}{Fabio Palomba}, \bibinfo{person}{Robert Feldt}, \bibinfo{person}{Rodrigo Spinola}, {and} \bibinfo{person}{David Lo}.} \bibinfo{year}{2026}\natexlab{a}.
\newblock \bibinfo{title}{The Hidden Methodological Layer: Uncovering Threats to Validity in {LLM}-Assisted Empirical Software Engineering}.
\newblock
\urldef\tempurl%
\url{https://github.com/secentervcu/LLM-Assisted-ESE-Threats-To-Validity-SLR}
\showURL{%
\tempurl}


\bibitem[Gomes et~al\mbox{.}(2026b)]%
        {gomes2026llmassistedESE}
\bibfield{author}{\bibinfo{person}{Victoria Gomes}, \bibinfo{person}{Delaney Selb}, \bibinfo{person}{Fabio Palomba}, \bibinfo{person}{Rodrigo Spinola}, {and} \bibinfo{person}{David Lo}.} \bibinfo{year}{2026}\natexlab{b}.
\newblock \bibinfo{title}{{LLM}-Assisted Empirical Software Engineering: Systematic Literature Review and Research Agenda}.
\newblock
\showeprint[arxiv]{2604.26192}~[cs.SE]
\urldef\tempurl%
\url{https://arxiv.org/abs/2604.26192}
\showURL{%
\tempurl}


\bibitem[Henderson et~al\mbox{.}(2018)]%
        {henderson2018deep}
\bibfield{author}{\bibinfo{person}{Peter Henderson}, \bibinfo{person}{Riashat Islam}, \bibinfo{person}{Philip Bachman}, \bibinfo{person}{Joelle Pineau}, \bibinfo{person}{Doina Precup}, {and} \bibinfo{person}{David Meger}.} \bibinfo{year}{2018}\natexlab{}.
\newblock \showarticletitle{Deep reinforcement learning that matters}. In \bibinfo{booktitle}{\emph{Proceedings of the Thirty-Second AAAI Conference on Artificial Intelligence and Thirtieth Innovative Applications of Artificial Intelligence Conference and Eighth AAAI Symposium on Educational Advances in Artificial Intelligence}} (New Orleans, Louisiana, USA) \emph{(\bibinfo{series}{AAAI'18/IAAI'18/EAAI'18})}. \bibinfo{publisher}{AAAI Press}, Article \bibinfo{articleno}{392}, \bibinfo{numpages}{8}~pages.
\newblock
\showISBNx{978-1-57735-800-8}


\bibitem[Hou et~al\mbox{.}(2024)]%
        {Hou2023}
\bibfield{author}{\bibinfo{person}{Xinyi Hou}, \bibinfo{person}{Yanjie Zhao}, \bibinfo{person}{Yue Liu}, \bibinfo{person}{Zhou Yang}, \bibinfo{person}{Kailong Wang}, \bibinfo{person}{Li Li}, \bibinfo{person}{Xiapu Luo}, \bibinfo{person}{David Lo}, \bibinfo{person}{John Grundy}, {and} \bibinfo{person}{Haoyu Wang}.} \bibinfo{year}{2024}\natexlab{}.
\newblock \showarticletitle{Large Language Models for Software Engineering: A Systematic Literature Review}.
\newblock \bibinfo{journal}{\emph{ACM Trans. Softw. Eng. Methodol.}} \bibinfo{volume}{33}, \bibinfo{number}{8}, Article \bibinfo{articleno}{220} (\bibinfo{date}{Dec.} \bibinfo{year}{2024}), \bibinfo{numpages}{79}~pages.
\newblock
\showISSN{1049-331X}
\href{https://doi.org/10.1145/3695988}{doi:\nolinkurl{10.1145/3695988}}


\bibitem[Jedlitschka et~al\mbox{.}(2008)]%
        {Jedlitschka2008}
\bibfield{author}{\bibinfo{person}{Andreas Jedlitschka}, \bibinfo{person}{Marcus Ciolkowski}, {and} \bibinfo{person}{Dietmar Pfahl}.} \bibinfo{year}{2008}\natexlab{}.
\newblock \showarticletitle{Reporting Experiments in Software Engineering}.
\newblock In \bibinfo{booktitle}{\emph{Guide to Advanced Empirical Software Engineering}}, \bibfield{editor}{\bibinfo{person}{Forrest Shull}, \bibinfo{person}{Janice Singer}, {and} \bibinfo{person}{Dag I.~K. Sj{\o}berg}} (Eds.). \bibinfo{publisher}{Springer London}, \bibinfo{address}{London}, \bibinfo{pages}{201--228}.
\newblock
\showISBNx{978-1-84800-044-5}
\href{https://doi.org/10.1007/978-1-84800-044-5_8}{doi:\nolinkurl{10.1007/978-1-84800-044-5_8}}


\bibitem[Ji et~al\mbox{.}(2023)]%
        {ji2023survey}
\bibfield{author}{\bibinfo{person}{Ziwei Ji}, \bibinfo{person}{Nayeon Lee}, \bibinfo{person}{Rita Frieske}, \bibinfo{person}{Tiezheng Yu}, \bibinfo{person}{Dan Su}, \bibinfo{person}{Yan Xu}, \bibinfo{person}{Etsuko Ishii}, \bibinfo{person}{Ye~Jin Bang}, \bibinfo{person}{Andrea Madotto}, {and} \bibinfo{person}{Pascale Fung}.} \bibinfo{year}{2023}\natexlab{}.
\newblock \showarticletitle{Survey of Hallucination in Natural Language Generation}.
\newblock \bibinfo{journal}{\emph{ACM Comput. Surv.}} \bibinfo{volume}{55}, \bibinfo{number}{12}, Article \bibinfo{articleno}{248} (\bibinfo{date}{March} \bibinfo{year}{2023}), \bibinfo{numpages}{38}~pages.
\newblock
\showISSN{0360-0300}
\href{https://doi.org/10.1145/3571730}{doi:\nolinkurl{10.1145/3571730}}


\bibitem[Kitchenham and Charters(2007)]%
        {kitchenham2007guidelines}
\bibfield{author}{\bibinfo{person}{Barbara Kitchenham} {and} \bibinfo{person}{Stuart Charters}.} \bibinfo{year}{2007}\natexlab{}.
\newblock \bibinfo{booktitle}{\emph{Guidelines for Performing Systematic Literature Reviews in Software Engineering}}.
\newblock \bibinfo{type}{EBSE Technical Report} EBSE-2007-01. \bibinfo{institution}{Keele University and Durham University}.
\newblock
\urldef\tempurl%
\url{https://www.elsevier.com/__data/promis_misc/525444systematicreviewsguide.pdf}
\showURL{%
\tempurl}


\bibitem[Kitchenham et~al\mbox{.}(2002)]%
        {Kitchenham2002preliminary}
\bibfield{author}{\bibinfo{person}{B.A. Kitchenham}, \bibinfo{person}{S.L. Pfleeger}, \bibinfo{person}{L.M. Pickard}, \bibinfo{person}{P.W. Jones}, \bibinfo{person}{D.C. Hoaglin}, \bibinfo{person}{K. El~Emam}, {and} \bibinfo{person}{J. Rosenberg}.} \bibinfo{year}{2002}\natexlab{}.
\newblock \showarticletitle{Preliminary guidelines for empirical research in software engineering}.
\newblock \bibinfo{journal}{\emph{IEEE Transactions on Software Engineering}} \bibinfo{volume}{28}, \bibinfo{number}{8} (\bibinfo{year}{2002}), \bibinfo{pages}{721--734}.
\newblock
\href{https://doi.org/10.1109/TSE.2002.1027796}{doi:\nolinkurl{10.1109/TSE.2002.1027796}}


\bibitem[Lin{\aa}ker et~al\mbox{.}(2015)]%
        {linaker2015guidelines}
\bibfield{author}{\bibinfo{person}{Johan Lin{\aa}ker}, \bibinfo{person}{Sardar~Muhammad Sulaman}, \bibinfo{person}{Rafael Maiani~de Mello}, {and} \bibinfo{person}{Martin H{\"o}st}.} \bibinfo{year}{2015}\natexlab{}.
\newblock \bibinfo{booktitle}{\emph{Guidelines for Conducting Surveys in Software Engineering}}.
\newblock \bibinfo{publisher}{Department of Computer Science, Lund University}.
\newblock


\bibitem[Mehrabi et~al\mbox{.}(2021)]%
        {mehrabi2021survey}
\bibfield{author}{\bibinfo{person}{Ninareh Mehrabi}, \bibinfo{person}{Fred Morstatter}, \bibinfo{person}{Nripsuta Saxena}, \bibinfo{person}{Kristina Lerman}, {and} \bibinfo{person}{Aram Galstyan}.} \bibinfo{year}{2021}\natexlab{}.
\newblock \showarticletitle{A Survey on Bias and Fairness in Machine Learning}.
\newblock \bibinfo{journal}{\emph{ACM Comput. Surv.}} \bibinfo{volume}{54}, \bibinfo{number}{6}, Article \bibinfo{articleno}{115} (\bibinfo{date}{July} \bibinfo{year}{2021}), \bibinfo{numpages}{35}~pages.
\newblock
\showISSN{0360-0300}
\href{https://doi.org/10.1145/3457607}{doi:\nolinkurl{10.1145/3457607}}


\bibitem[OpenAI et~al\mbox{.}(2024)]%
        {openai2024gpt4}
\bibfield{author}{\bibinfo{person}{OpenAI}, \bibinfo{person}{Josh Achiam}, \bibinfo{person}{Steven Adler}, \bibinfo{person}{Sandhini Agarwal}, \bibinfo{person}{Lama Ahmad}, \bibinfo{person}{Ilge Akkaya}, \bibinfo{person}{Florencia~Leoni Aleman}, \bibinfo{person}{Diogo Almeida}, \bibinfo{person}{Janko Altenschmidt}, \bibinfo{person}{Sam Altman}, \bibinfo{person}{Shyamal Anadkat}, \bibinfo{person}{Red Avila}, \bibinfo{person}{Igor Babuschkin}, \bibinfo{person}{Suchir Balaji}, \bibinfo{person}{Valerie Balcom}, \bibinfo{person}{Paul Baltescu}, \bibinfo{person}{Haiming Bao}, \bibinfo{person}{Mohammad Bavarian}, \bibinfo{person}{Jeff Belgum}, \bibinfo{person}{Irwan Bello}, \bibinfo{person}{Jake Berdine}, \bibinfo{person}{Gabriel Bernadett-Shapiro}, \bibinfo{person}{Christopher Berner}, \bibinfo{person}{Lenny Bogdonoff}, \bibinfo{person}{Oleg Boiko}, \bibinfo{person}{Madelaine Boyd}, \bibinfo{person}{Anna-Luisa Brakman}, \bibinfo{person}{Greg Brockman}, \bibinfo{person}{Tim Brooks}, \bibinfo{person}{Miles Brundage},
  \bibinfo{person}{Kevin Button}, \bibinfo{person}{Trevor Cai}, \bibinfo{person}{Rosie Campbell}, \bibinfo{person}{Andrew Cann}, \bibinfo{person}{Brittany Carey}, \bibinfo{person}{Chelsea Carlson}, \bibinfo{person}{Rory Carmichael}, \bibinfo{person}{Brooke Chan}, \bibinfo{person}{Che Chang}, \bibinfo{person}{Fotis Chantzis}, \bibinfo{person}{Derek Chen}, \bibinfo{person}{Sully Chen}, \bibinfo{person}{Ruby Chen}, \bibinfo{person}{Jason Chen}, \bibinfo{person}{Mark Chen}, \bibinfo{person}{Ben Chess}, \bibinfo{person}{Chester Cho}, \bibinfo{person}{Casey Chu}, \bibinfo{person}{Hyung~Won Chung}, \bibinfo{person}{Dave Cummings}, \bibinfo{person}{Jeremiah Currier}, \bibinfo{person}{Yunxing Dai}, \bibinfo{person}{Cory Decareaux}, \bibinfo{person}{Thomas Degry}, \bibinfo{person}{Noah Deutsch}, \bibinfo{person}{Damien Deville}, \bibinfo{person}{Arka Dhar}, \bibinfo{person}{David Dohan}, \bibinfo{person}{Steve Dowling}, \bibinfo{person}{Sheila Dunning}, \bibinfo{person}{Adrien Ecoffet}, \bibinfo{person}{Atty Eleti},
  \bibinfo{person}{Tyna Eloundou}, \bibinfo{person}{David Farhi}, \bibinfo{person}{Liam Fedus}, \bibinfo{person}{Niko Felix}, \bibinfo{person}{Simón~Posada Fishman}, \bibinfo{person}{Juston Forte}, \bibinfo{person}{Isabella Fulford}, \bibinfo{person}{Leo Gao}, \bibinfo{person}{Elie Georges}, \bibinfo{person}{Christian Gibson}, \bibinfo{person}{Vik Goel}, \bibinfo{person}{Tarun Gogineni}, \bibinfo{person}{Gabriel Goh}, \bibinfo{person}{Rapha Gontijo-Lopes}, \bibinfo{person}{Jonathan Gordon}, \bibinfo{person}{Morgan Grafstein}, \bibinfo{person}{Scott Gray}, \bibinfo{person}{Ryan Greene}, \bibinfo{person}{Joshua Gross}, \bibinfo{person}{Shixiang~Shane Gu}, \bibinfo{person}{Yufei Guo}, \bibinfo{person}{Chris Hallacy}, \bibinfo{person}{Jesse Han}, \bibinfo{person}{Jeff Harris}, \bibinfo{person}{Yuchen He}, \bibinfo{person}{Mike Heaton}, \bibinfo{person}{Johannes Heidecke}, \bibinfo{person}{Chris Hesse}, \bibinfo{person}{Alan Hickey}, \bibinfo{person}{Wade Hickey}, \bibinfo{person}{Peter Hoeschele},
  \bibinfo{person}{Brandon Houghton}, \bibinfo{person}{Kenny Hsu}, \bibinfo{person}{Shengli Hu}, \bibinfo{person}{Xin Hu}, \bibinfo{person}{Joost Huizinga}, \bibinfo{person}{Shantanu Jain}, \bibinfo{person}{Shawn Jain}, \bibinfo{person}{Joanne Jang}, \bibinfo{person}{Angela Jiang}, \bibinfo{person}{Roger Jiang}, \bibinfo{person}{Haozhun Jin}, \bibinfo{person}{Denny Jin}, \bibinfo{person}{Shino Jomoto}, \bibinfo{person}{Billie Jonn}, \bibinfo{person}{Heewoo Jun}, \bibinfo{person}{Tomer Kaftan}, \bibinfo{person}{Łukasz Kaiser}, \bibinfo{person}{Ali Kamali}, \bibinfo{person}{Ingmar Kanitscheider}, \bibinfo{person}{Nitish~Shirish Keskar}, \bibinfo{person}{Tabarak Khan}, \bibinfo{person}{Logan Kilpatrick}, \bibinfo{person}{Jong~Wook Kim}, \bibinfo{person}{Christina Kim}, \bibinfo{person}{Yongjik Kim}, \bibinfo{person}{Jan~Hendrik Kirchner}, \bibinfo{person}{Jamie Kiros}, \bibinfo{person}{Matt Knight}, \bibinfo{person}{Daniel Kokotajlo}, \bibinfo{person}{Łukasz Kondraciuk}, \bibinfo{person}{Andrew Kondrich},
  \bibinfo{person}{Aris Konstantinidis}, \bibinfo{person}{Kyle Kosic}, \bibinfo{person}{Gretchen Krueger}, \bibinfo{person}{Vishal Kuo}, \bibinfo{person}{Michael Lampe}, \bibinfo{person}{Ikai Lan}, \bibinfo{person}{Teddy Lee}, \bibinfo{person}{Jan Leike}, \bibinfo{person}{Jade Leung}, \bibinfo{person}{Daniel Levy}, \bibinfo{person}{Chak~Ming Li}, \bibinfo{person}{Rachel Lim}, \bibinfo{person}{Molly Lin}, \bibinfo{person}{Stephanie Lin}, \bibinfo{person}{Mateusz Litwin}, \bibinfo{person}{Theresa Lopez}, \bibinfo{person}{Ryan Lowe}, \bibinfo{person}{Patricia Lue}, \bibinfo{person}{Anna Makanju}, \bibinfo{person}{Kim Malfacini}, \bibinfo{person}{Sam Manning}, \bibinfo{person}{Todor Markov}, \bibinfo{person}{Yaniv Markovski}, \bibinfo{person}{Bianca Martin}, \bibinfo{person}{Katie Mayer}, \bibinfo{person}{Andrew Mayne}, \bibinfo{person}{Bob McGrew}, \bibinfo{person}{Scott~Mayer McKinney}, \bibinfo{person}{Christine McLeavey}, \bibinfo{person}{Paul McMillan}, \bibinfo{person}{Jake McNeil}, \bibinfo{person}{David
  Medina}, \bibinfo{person}{Aalok Mehta}, \bibinfo{person}{Jacob Menick}, \bibinfo{person}{Luke Metz}, \bibinfo{person}{Andrey Mishchenko}, \bibinfo{person}{Pamela Mishkin}, \bibinfo{person}{Vinnie Monaco}, \bibinfo{person}{Evan Morikawa}, \bibinfo{person}{Daniel Mossing}, \bibinfo{person}{Tong Mu}, \bibinfo{person}{Mira Murati}, \bibinfo{person}{Oleg Murk}, \bibinfo{person}{David Mély}, \bibinfo{person}{Ashvin Nair}, \bibinfo{person}{Reiichiro Nakano}, \bibinfo{person}{Rajeev Nayak}, \bibinfo{person}{Arvind Neelakantan}, \bibinfo{person}{Richard Ngo}, \bibinfo{person}{Hyeonwoo Noh}, \bibinfo{person}{Long Ouyang}, \bibinfo{person}{Cullen O'Keefe}, \bibinfo{person}{Jakub Pachocki}, \bibinfo{person}{Alex Paino}, \bibinfo{person}{Joe Palermo}, \bibinfo{person}{Ashley Pantuliano}, \bibinfo{person}{Giambattista Parascandolo}, \bibinfo{person}{Joel Parish}, \bibinfo{person}{Emy Parparita}, \bibinfo{person}{Alex Passos}, \bibinfo{person}{Mikhail Pavlov}, \bibinfo{person}{Andrew Peng}, \bibinfo{person}{Adam
  Perelman}, \bibinfo{person}{Filipe de Avila Belbute~Peres}, \bibinfo{person}{Michael Petrov}, \bibinfo{person}{Henrique~Ponde de Oliveira~Pinto}, \bibinfo{person}{Michael}, \bibinfo{person}{Pokorny}, \bibinfo{person}{Michelle Pokrass}, \bibinfo{person}{Vitchyr~H. Pong}, \bibinfo{person}{Tolly Powell}, \bibinfo{person}{Alethea Power}, \bibinfo{person}{Boris Power}, \bibinfo{person}{Elizabeth Proehl}, \bibinfo{person}{Raul Puri}, \bibinfo{person}{Alec Radford}, \bibinfo{person}{Jack Rae}, \bibinfo{person}{Aditya Ramesh}, \bibinfo{person}{Cameron Raymond}, \bibinfo{person}{Francis Real}, \bibinfo{person}{Kendra Rimbach}, \bibinfo{person}{Carl Ross}, \bibinfo{person}{Bob Rotsted}, \bibinfo{person}{Henri Roussez}, \bibinfo{person}{Nick Ryder}, \bibinfo{person}{Mario Saltarelli}, \bibinfo{person}{Ted Sanders}, \bibinfo{person}{Shibani Santurkar}, \bibinfo{person}{Girish Sastry}, \bibinfo{person}{Heather Schmidt}, \bibinfo{person}{David Schnurr}, \bibinfo{person}{John Schulman}, \bibinfo{person}{Daniel Selsam},
  \bibinfo{person}{Kyla Sheppard}, \bibinfo{person}{Toki Sherbakov}, \bibinfo{person}{Jessica Shieh}, \bibinfo{person}{Sarah Shoker}, \bibinfo{person}{Pranav Shyam}, \bibinfo{person}{Szymon Sidor}, \bibinfo{person}{Eric Sigler}, \bibinfo{person}{Maddie Simens}, \bibinfo{person}{Jordan Sitkin}, \bibinfo{person}{Katarina Slama}, \bibinfo{person}{Ian Sohl}, \bibinfo{person}{Benjamin Sokolowsky}, \bibinfo{person}{Yang Song}, \bibinfo{person}{Natalie Staudacher}, \bibinfo{person}{Felipe~Petroski Such}, \bibinfo{person}{Natalie Summers}, \bibinfo{person}{Ilya Sutskever}, \bibinfo{person}{Jie Tang}, \bibinfo{person}{Nikolas Tezak}, \bibinfo{person}{Madeleine~B. Thompson}, \bibinfo{person}{Phil Tillet}, \bibinfo{person}{Amin Tootoonchian}, \bibinfo{person}{Elizabeth Tseng}, \bibinfo{person}{Preston Tuggle}, \bibinfo{person}{Nick Turley}, \bibinfo{person}{Jerry Tworek}, \bibinfo{person}{Juan Felipe~Cerón Uribe}, \bibinfo{person}{Andrea Vallone}, \bibinfo{person}{Arun Vijayvergiya}, \bibinfo{person}{Chelsea Voss},
  \bibinfo{person}{Carroll Wainwright}, \bibinfo{person}{Justin~Jay Wang}, \bibinfo{person}{Alvin Wang}, \bibinfo{person}{Ben Wang}, \bibinfo{person}{Jonathan Ward}, \bibinfo{person}{Jason Wei}, \bibinfo{person}{CJ Weinmann}, \bibinfo{person}{Akila Welihinda}, \bibinfo{person}{Peter Welinder}, \bibinfo{person}{Jiayi Weng}, \bibinfo{person}{Lilian Weng}, \bibinfo{person}{Matt Wiethoff}, \bibinfo{person}{Dave Willner}, \bibinfo{person}{Clemens Winter}, \bibinfo{person}{Samuel Wolrich}, \bibinfo{person}{Hannah Wong}, \bibinfo{person}{Lauren Workman}, \bibinfo{person}{Sherwin Wu}, \bibinfo{person}{Jeff Wu}, \bibinfo{person}{Michael Wu}, \bibinfo{person}{Kai Xiao}, \bibinfo{person}{Tao Xu}, \bibinfo{person}{Sarah Yoo}, \bibinfo{person}{Kevin Yu}, \bibinfo{person}{Qiming Yuan}, \bibinfo{person}{Wojciech Zaremba}, \bibinfo{person}{Rowan Zellers}, \bibinfo{person}{Chong Zhang}, \bibinfo{person}{Marvin Zhang}, \bibinfo{person}{Shengjia Zhao}, \bibinfo{person}{Tianhao Zheng}, \bibinfo{person}{Juntang Zhuang},
  \bibinfo{person}{William Zhuk}, {and} \bibinfo{person}{Barret Zoph}.} \bibinfo{year}{2024}\natexlab{}.
\newblock \bibinfo{title}{GPT-4 Technical Report}.
\newblock
\showeprint[arxiv]{2303.08774}~[cs.CL]
\urldef\tempurl%
\url{https://arxiv.org/abs/2303.08774}
\showURL{%
\tempurl}


\bibitem[Reynolds and McDonell(2021)]%
        {reynolds2021prompt}
\bibfield{author}{\bibinfo{person}{Laria Reynolds} {and} \bibinfo{person}{Kyle McDonell}.} \bibinfo{year}{2021}\natexlab{}.
\newblock \showarticletitle{Prompt Programming for Large Language Models: Beyond the Few-Shot Paradigm}. In \bibinfo{booktitle}{\emph{Extended Abstracts of the 2021 CHI Conference on Human Factors in Computing Systems}} (Yokohama, Japan) \emph{(\bibinfo{series}{CHI EA '21})}. \bibinfo{publisher}{Association for Computing Machinery}, \bibinfo{address}{New York, NY, USA}, Article \bibinfo{articleno}{314}, \bibinfo{numpages}{7}~pages.
\newblock
\showISBNx{9781450380959}
\href{https://doi.org/10.1145/3411763.3451760}{doi:\nolinkurl{10.1145/3411763.3451760}}


\bibitem[Runeson and H\"{o}st(2009)]%
        {Runeson2009guidelines}
\bibfield{author}{\bibinfo{person}{Per Runeson} {and} \bibinfo{person}{Martin H\"{o}st}.} \bibinfo{year}{2009}\natexlab{}.
\newblock \showarticletitle{Guidelines for conducting and reporting case study research in software engineering}.
\newblock \bibinfo{journal}{\emph{Empirical Softw. Engg.}} \bibinfo{volume}{14}, \bibinfo{number}{2} (\bibinfo{date}{April} \bibinfo{year}{2009}), \bibinfo{pages}{131–164}.
\newblock
\showISSN{1382-3256}
\href{https://doi.org/10.1007/s10664-008-9102-8}{doi:\nolinkurl{10.1007/s10664-008-9102-8}}


\bibitem[Sallou et~al\mbox{.}(2024)]%
        {Sallou2024breaking}
\bibfield{author}{\bibinfo{person}{June Sallou}, \bibinfo{person}{Thomas Durieux}, {and} \bibinfo{person}{Annibale Panichella}.} \bibinfo{year}{2024}\natexlab{}.
\newblock \showarticletitle{Breaking the Silence: the Threats of Using {LLM}s in Software Engineering}. In \bibinfo{booktitle}{\emph{Proceedings of the 2024 ACM/IEEE 44th International Conference on Software Engineering: New Ideas and Emerging Results}} (Lisbon, Portugal) \emph{(\bibinfo{series}{ICSE-NIER'24})}. \bibinfo{publisher}{Association for Computing Machinery}, \bibinfo{address}{New York, NY, USA}, \bibinfo{pages}{102–106}.
\newblock
\showISBNx{9798400705007}
\href{https://doi.org/10.1145/3639476.3639764}{doi:\nolinkurl{10.1145/3639476.3639764}}


\bibitem[Shneiderman(2020)]%
        {Shneiderman2020human}
\bibfield{author}{\bibinfo{person}{Ben Shneiderman}.} \bibinfo{year}{2020}\natexlab{}.
\newblock \showarticletitle{Human-Centered Artificial Intelligence: Reliable, Safe \& Trustworthy}.
\newblock \bibinfo{journal}{\emph{International Journal of Human–Computer Interaction}} \bibinfo{volume}{36}, \bibinfo{number}{6} (\bibinfo{year}{2020}), \bibinfo{pages}{495--504}.
\newblock
\href{https://doi.org/10.1080/10447318.2020.1741118}{doi:\nolinkurl{10.1080/10447318.2020.1741118}}


\bibitem[Shull et~al\mbox{.}(2008)]%
        {Shull2008}
\bibfield{author}{\bibinfo{person}{Forrest Shull}, \bibinfo{person}{Janice Singer}, {and} \bibinfo{person}{Dag I.~K. Sj{\o}berg}.} \bibinfo{year}{2008}\natexlab{}.
\newblock \bibinfo{booktitle}{\emph{Guide to Advanced Empirical Software Engineering}}.
\newblock \bibinfo{publisher}{Springer}.
\newblock
\href{https://doi.org/10.1007/978-1-84800-044-5}{doi:\nolinkurl{10.1007/978-1-84800-044-5}}


\bibitem[Teixeira et~al\mbox{.}(2025)]%
        {Muller2023genaiqual}
\bibfield{author}{\bibinfo{person}{Maria~Couto Teixeira}, \bibinfo{person}{Marisa Tschopp}, {and} \bibinfo{person}{Anna Jobin}.} \bibinfo{year}{2025}\natexlab{}.
\newblock \bibinfo{title}{Generative Artificial Intelligence in Qualitative Research Methods: Between Hype and Risks?}
\newblock
\showeprint[arxiv]{2511.08461}~[cs.CY]
\urldef\tempurl%
\url{https://arxiv.org/abs/2511.08461}
\showURL{%
\tempurl}


\bibitem[Wagner et~al\mbox{.}(2025)]%
        {Wagner2025guidelines}
\bibfield{author}{\bibinfo{person}{Stefan Wagner}, \bibinfo{person}{Marvin Mu\~{n}oz Bar\'{o}n}, \bibinfo{person}{Davide Falessi}, {and} \bibinfo{person}{Sebastian Baltes}.} \bibinfo{year}{2025}\natexlab{}.
\newblock \showarticletitle{Towards Evaluation Guidelines for Empirical Studies Involving {LLM}s}. In \bibinfo{booktitle}{\emph{2025 IEEE/ACM International Workshop on Methodological Issues with Empirical Studies in Software Engineering (WSESE)}} (Ottawa, ON, Canada). \bibinfo{publisher}{IEEE Press}, \bibinfo{pages}{24–27}.
\newblock
\href{https://doi.org/10.1109/WSESE66602.2025.00011}{doi:\nolinkurl{10.1109/WSESE66602.2025.00011}}


\bibitem[Wei et~al\mbox{.}(2022)]%
        {Wei2022chain}
\bibfield{author}{\bibinfo{person}{Jason Wei}, \bibinfo{person}{Xuezhi Wang}, \bibinfo{person}{Dale Schuurmans}, \bibinfo{person}{Maarten Bosma}, \bibinfo{person}{Brian Ichter}, \bibinfo{person}{Fei Xia}, \bibinfo{person}{Ed~H. Chi}, \bibinfo{person}{Quoc~V. Le}, {and} \bibinfo{person}{Denny Zhou}.} \bibinfo{year}{2022}\natexlab{}.
\newblock \showarticletitle{Chain-of-thought prompting elicits reasoning in large language models}. In \bibinfo{booktitle}{\emph{Proceedings of the 36th International Conference on Neural Information Processing Systems}} (New Orleans, LA, USA) \emph{(\bibinfo{series}{NIPS '22})}. \bibinfo{publisher}{Curran Associates Inc.}, \bibinfo{address}{Red Hook, NY, USA}, Article \bibinfo{articleno}{1800}, \bibinfo{numpages}{14}~pages.
\newblock
\showISBNx{9781713871088}


\bibitem[Weidinger et~al\mbox{.}(2021)]%
        {Weidinger2022}
\bibfield{author}{\bibinfo{person}{Laura Weidinger}, \bibinfo{person}{John Mellor}, \bibinfo{person}{Maribeth Rauh}, \bibinfo{person}{Conor Griffin}, \bibinfo{person}{Jonathan Uesato}, \bibinfo{person}{Po-Sen Huang}, \bibinfo{person}{Myra Cheng}, \bibinfo{person}{Mia Glaese}, \bibinfo{person}{Borja Balle}, \bibinfo{person}{Atoosa Kasirzadeh}, \bibinfo{person}{Zac Kenton}, \bibinfo{person}{Sasha Brown}, \bibinfo{person}{Will Hawkins}, \bibinfo{person}{Tom Stepleton}, \bibinfo{person}{Courtney Biles}, \bibinfo{person}{Abeba Birhane}, \bibinfo{person}{Julia Haas}, \bibinfo{person}{Laura Rimell}, \bibinfo{person}{Lisa~Anne Hendricks}, \bibinfo{person}{William Isaac}, \bibinfo{person}{Sean Legassick}, \bibinfo{person}{Geoffrey Irving}, {and} \bibinfo{person}{Iason Gabriel}.} \bibinfo{year}{2021}\natexlab{}.
\newblock \bibinfo{title}{Ethical and social risks of harm from Language Models}.
\newblock
\showeprint[arxiv]{2112.04359}~[cs.CL]
\urldef\tempurl%
\url{https://arxiv.org/abs/2112.04359}
\showURL{%
\tempurl}


\bibitem[Wohlin et~al\mbox{.}(2024)]%
        {ExperimentationSEWohlin}
\bibfield{author}{\bibinfo{person}{Claes Wohlin}, \bibinfo{person}{Per Runeson}, \bibinfo{person}{Martin Hst}, \bibinfo{person}{Magnus~C. Ohlsson}, \bibinfo{person}{Bjrn Regnell}, {and} \bibinfo{person}{Anders Wessln}.} \bibinfo{year}{2024}\natexlab{}.
\newblock \bibinfo{booktitle}{\emph{Experimentation in Software Engineering}}.
\newblock \bibinfo{publisher}{Springer Publishing Company, Incorporated}.
\newblock
\showISBNx{9783662693063}
\href{https://doi.org/10.1007/978-3-662-69306-3}{doi:\nolinkurl{10.1007/978-3-662-69306-3}}


\bibitem[Zhang et~al\mbox{.}(2018)]%
        {Zhang2018empirical}
\bibfield{author}{\bibinfo{person}{Li Zhang}, \bibinfo{person}{Jia-Hao Tian}, \bibinfo{person}{Jing Jiang}, \bibinfo{person}{Yi-Jun Liu}, \bibinfo{person}{Meng-Yuan Pu}, {and} \bibinfo{person}{Tao Yue}.} \bibinfo{year}{2018}\natexlab{}.
\newblock \showarticletitle{Empirical Research in Software Engineering-A Literature Survey}.
\newblock \bibinfo{journal}{\emph{Journal of Computer Science and Technology}} \bibinfo{volume}{33}, \bibinfo{number}{5} (\bibinfo{year}{2018}), \bibinfo{pages}{876--899}.
\newblock
\showISSN{1000-9000(Print) /1860-4749(Online)}
\href{https://doi.org/10.1007/s11390-018-1864-x}{doi:\nolinkurl{10.1007/s11390-018-1864-x}}


\bibitem[Zhang et~al\mbox{.}(2025)]%
        {Zhang2025}
\bibfield{author}{\bibinfo{person}{Yanbo Zhang}, \bibinfo{person}{Sumeer~A. Khan}, \bibinfo{person}{Adnan Mahmud}, \bibinfo{person}{Huck Yang}, \bibinfo{person}{Alexander Lavin}, \bibinfo{person}{Michael Levin}, \bibinfo{person}{Jeremy Frey}, \bibinfo{person}{Jared Dunnmon}, \bibinfo{person}{James Evans}, \bibinfo{person}{Alan Bundy}, \bibinfo{person}{Saso Dzeroski}, \bibinfo{person}{Jesper Tegner}, {and} \bibinfo{person}{Hector Zenil}.} \bibinfo{year}{2025}\natexlab{}.
\newblock \showarticletitle{Exploring the role of large language models in the scientific method: from hypothesis to discovery}.
\newblock \bibinfo{journal}{\emph{npj Artificial Intelligence}} \bibinfo{volume}{1}, \bibinfo{number}{1} (\bibinfo{year}{2025}), \bibinfo{pages}{14}.
\newblock
\showISSN{3005-1460}
\href{https://doi.org/10.1038/s44387-025-00019-5}{doi:\nolinkurl{10.1038/s44387-025-00019-5}}


\bibitem[Zhao et~al\mbox{.}(2021)]%
        {zhao2021calibrate}
\bibfield{author}{\bibinfo{person}{Zihao Zhao}, \bibinfo{person}{Eric Wallace}, \bibinfo{person}{Shi Feng}, \bibinfo{person}{Dan Klein}, {and} \bibinfo{person}{Sameer Singh}.} \bibinfo{year}{2021}\natexlab{}.
\newblock \showarticletitle{Calibrate Before Use: Improving Few-shot Performance of Language Models}. In \bibinfo{booktitle}{\emph{Proceedings of the 38th International Conference on Machine Learning}} \emph{(\bibinfo{series}{Proceedings of Machine Learning Research}, Vol.~\bibinfo{volume}{139})}, \bibfield{editor}{\bibinfo{person}{Marina Meila} {and} \bibinfo{person}{Tong Zhang}} (Eds.). \bibinfo{publisher}{PMLR}, \bibinfo{pages}{12697--12706}.
\newblock
\urldef\tempurl%
\url{https://proceedings.mlr.press/v139/zhao21c.html}
\showURL{%
\tempurl}


\end{thebibliography}

\end{document}